\documentclass[12pt]{article} 

\RequirePackage[T1]{fontenc}

\usepackage[height=8.85in,width=6.45in]{geometry}
\usepackage[utf8]{inputenc}
\usepackage{amsmath}
\usepackage{amssymb}
\usepackage{mathtools}
\numberwithin{equation}{section}
\usepackage{slashed}
\usepackage{braket}
\usepackage[svgnames]{xcolor}
\usepackage[colorlinks,citecolor=DarkGreen,linkcolor=FireBrick]{hyperref}
\usepackage{cite}
\usepackage{graphicx}
\usepackage{tikz}
\usepackage{tikz-cd}
\usepackage{times}
\usepackage{courier}
\usepackage{bm}
\usepackage{subfig}
\usepackage{xcolor}
\usepackage{mdframed}
\usepackage{datetime}
\usepackage{dsfont}

\usepackage{colortbl}
\definecolor{dgreen}{rgb}{0, 0.55, 0}
\definecolor{llightyellow}{rgb}{1.0, 0.95, 0.7}
\definecolor{llightblue}{rgb}{0.7, 0.9, 1.0}
\definecolor{llightpink}{rgb}{1.0, 0.85, 0.95}
\definecolor{llightgreen}{rgb}{0.7, 1.0, 0.4}
\colorlet{lightyellow}{llightyellow!50!white}
\colorlet{lightblue}{llightblue!50!white}
\colorlet{lightgreen}{llightgreen!50!white}
\colorlet{lightpink}{llightpink!50!white}

\renewenvironment{figure}[1][]{
  \begin{originalfigure}[#1]
    \begin{mdframed}[linecolor=black!0,backgroundcolor=black!1]
}{
    \end{mdframed}
  \end{originalfigure}
}

\newcommand{\fu}{\mathfrak{u}}
\newcommand{\ft}{\mathfrak{t}}

\def\CH{{\mathcal H}}

\newcommand{\Z}{\mathbb{Z}}
\newcommand{\Tr}{\mathrm{Tr}}

\newcommand{\mS}{\mathsf{S}}
\newcommand{\mT}{\mathsf{T}}
\newcommand{\mU}{\mathsf{U}}
\newcommand{\mV}{\mathsf{V}}
\newcommand{\Kitaev}{\mathsf{Kitaev}}
\newcommand{\Arf}{\text{Arf}}
\newcommand{\twine}{symmetry }

\begin{document}
\begin{titlepage}

\begin{flushright}
\end{flushright}

\vskip 3cm

\begin{center}	
	{\Large \bfseries Boson-fermion duality  with subsystem symmetry}	
	\vskip 1cm

	Weiguang Cao$^{1,2}$, Masahito Yamazaki$^1$, and Yunqin Zheng$^{1,3}$ 	
	\vskip 1cm
	
	\begin{tabular}{ll}
		$^1$&Kavli Institute for the Physics and Mathematics of the Universe, \\
		& University of Tokyo,  Kashiwa, Chiba 277-8583, Japan\\
		$^2$&Department of Physics, Graduate School of Science,\\
		& University of Tokyo, Tokyo 113-0033, Japan\\
		$^3$&Institute for Solid State Physics, \\
		&University of Tokyo,  Kashiwa, Chiba 277-8581, Japan\\
	\end{tabular}
	\vskip 1cm
	
\end{center}

\noindent
We explore an exact duality in $(2+1)$d between the fermionization of a bosonic theory with a $\mathbb{Z}_2$ subsystem symmetry and a fermionic theory with a $\mathbb{Z}_2$ subsystem fermion parity symmetry. A typical example is the duality between the fermionization of the plaquette Ising model and the plaquette fermion model. We first revisit the standard boson-fermion duality in $(1+1)$d with a $\mathbb{Z}_2$ 0-form symmetry, presenting in a way generalizable to $(2+1)$d.  We proceed to $(2+1)$d with a $\mathbb{Z}_2$ subsystem symmetry and establish the exact duality on the lattice by using the generalized Jordan-Wigner map, with a careful discussion on the mapping of the twist and symmetry sectors. This motivates us to introduce the subsystem Arf invariant, which exhibits a foliation structure. 

\end{titlepage}

\setcounter{tocdepth}{3}
\tableofcontents

\section{Introduction and summary}

\paragraph{Subsystem symmetries:}
Systems in fracton topological ordered phases are exotic class of gapped systems that have attracted much attention during the past decade \cite{Haah:2011drr, Haah:2013eia, Vijay:2016phm,Ma:2017aog,Shirley:2017suz,Shirley:2018nhn,Shirley:2018hkm,Slagle:2020ugk, Tian:2018opt, Shen:2021rct}. These systems share many interesting features, including (1) exponentially growing ground state degeneracy with respect to the system size \cite{Vijay:2016phm, Haah:2011drr, Haah:2013eia}, (2) restricted mobility of the excitations above the ground states \cite{Vijay:2016phm}, and (3) large subleading corrections to the entanglement entropy\cite{2018PhRvB..97l5102H, 2018PhRvB..97l5101M}. More details can be found in the reviews \cite{Nandkishore:2018sel,Pretko:2020cko}.
A significant portion of the fracton topological ordered systems can be obtained from gauging subsystem symmetries \cite{Vijay:2016phm, Shirley:2018vtc,Ibieta-Jimenez:2019zsf}. Thus it is equally interesting and important to study subsystem-symmetric theories. See \cite{You:2018oai, PhysRevB.98.235121, 2018arXiv180504097D, Kubica:2018lhn, 2019PhRvX...9b1010P, Shirley:2018vtc, 2019PhRvL.122n0506W} for early works on subsystem symmetries in exactly solvable models \footnote{Also see \cite{Batista:2004sc,Nussinov:2006iva,Nussinov:2009zz} where the subsystem symmetry is termed as "d-dimensional Gauge-Like Symmetries"}. 

The subsystem symmetry is one of many recent generalizations of the notion of global symmetries. In contrast to an ordinary 0-form global symmetry, where the symmetry generator is supported on the entire spatial manifold and acts on the entire Hilbert space, the generator of a subsystem symmetry is supported on a spatial submanifold and acts only on a subspace of the Hilbert space. Also in contrast to the higher-form symmetry \cite{Gaiotto:2014kfa}, the symmetry generator/defect for a subsystem symmetry is not topological under arbitrary deformations: in certain situations, it can be deformed only within a submanifold.  Despite its exoticness, a subsystem global symmetry shares many features with ordinary (0-form or higher-form) global symmetries. To name a few: 
    (1)  It enforces a selection rule of the correlation functions \cite{Gorantla:2021bda};
    (2) It can be spontaneously broken\cite{Qi:2020jrf,Distler:2021qzc,Rayhaun:2021ocs};
    (3) It can be anomalous and the anomaly can be canceled by the anomaly-inflow mechanism\cite{Burnell:2021reh};
    (4) As it exists throughout the RG flow, one can use it (and its anomaly) to constrain the dynamics in the long-distance limit \cite{Seiberg:2020wsg, Seiberg:2020bhn, Seiberg:2020cxy,SanMiguel:2020xxh,You:2020ykc}. 
In this work, we focus on one of the simplest possibilities: $\Z_2$ subsystem symmetry in $(2+1)$d. 

\paragraph{Boson-fermion duality:}
Boson-fermion duality among $(1+1)$d quantum field theories (QFTs) is an old subject, which was extensively  explored in the 70's and 80's \cite{Coleman:1974bu,Haldane_1981, Witten:1983ar}. The subject was revived recently due to the renewed understanding of discrete / higher-form symmetries, their gauging, and their anomalies \cite{Hsieh:2020uwb, Fukusumi:2021zme,Ebisu:2021acm, Karch:2019lnn,Ji:2019ugf,Gaiotto:2014kfa}. For example, it has been realized that the boson-fermion duality between the critical Ising model and the free Majorana fermion can be generalized to the duality between an arbitrary $(1+1)$d QFT with a non-anomalous $\Z_2$ global symmetry and a fermionic theory with a $\Z_2^F$ fermion parity symmetry. Moreover, the role of global-symmetry sectors (including $\Z_2$ twist sectors, and NS, R spin structures) has been emphasized by coupling to discrete (background) gauge fields. Another interesting line of development is the study of boson-fermion duality in $(2+1)$d, where a Dirac fermion was conjectured to be dual to the gauged Wilson-Fisher model suitably coupled to a Chern Simons term \cite{Seiberg:2016gmd, Turner:2019wnh, Karch:2016sxi, Hsin:2016blu, Aharony:2016jvv}. A huge duality web was then proposed based on this seed duality. 

Despite the exciting developments in recent years, some puzzles remain. Most of the recent developments mentioned above are about the \emph{infrared} dualities: the bosonic and fermionic theories are considered to be the dual to each other only in the long-distance limit. On the other hand, the prototypical example of boson-fermion duality in $(1+1)$d was defined via an exact Jordan-Wigner map. The Jordan-Wigner map defines the fermionic operator in terms of non-local expressions of the spin (bosonic) operators, and the Hamiltonian in terms of the spin operators is exactly rewritten as the Hamiltonian in terms of fermionic operators. Hence it defines an exact boson-fermion duality (as opposed to an infrared duality).  It was considered to be hard, however, to generalize the Jordan-Wigner map to higher dimensions. This is because when defining the fermion operators in terms of spin operators, one needs to assign tails of the spins to ensure the anti-commuting statistics of the fermions. However, assigning tails would generically break the rotation symmetry of the model. If we demand that the dual pairs of theories admit QFT descriptions that preserve the rotation (or even Lorentz) symmetry, it appears hard to generalize the exact boson-fermion duality to $(2+1)$d or higher, as the Jordan-Wigner map does not preserve the spatial rotation symmetry. See \cite{Chen:2017fvr,Chen:2018nog, Chen:2019wlx} for the approaches without explicitly specifying the tails.\footnote{See also \cite{Li:2021hvz} for a higher-dimensional Jordan-Wigner transformation keeping the spatial symmetry manifest, but with auxiliary Majorana Fermions and with constraints on the number of sites and \cite{Nussinov:2012nz} for discussion on fermionization of specific models with subsystem symmetry}

\paragraph{Boson-fermion duality of subsystem-symmetric systems:}
Subsystem-symmetric models open a natural window to generalizing the Jordan-Wigner map to higher dimensions. This is because subsystem-symmetric systems are generically defined on the lattice, and generically do not admit a continuum QFT description with Lorentz symmetry in the long-distance limit. See \cite{Wang:2019aiq, Wang:2019cbj, Seiberg:2020bhn,Seiberg:2020cxy, Seiberg:2020wsg, Gorantla:2020xap, Gorantla:2020jpy, Rudelius:2020kta, Gorantla:2021svj, Gorantla:2021bda,Gorantla:2022eem} for recent developments on the QFT descriptions with only discrete spatial rotation  symmetries. Hence the problem mentioned in the previous paragraph is no longer a problem in this setting. Indeed, we will see in the following sections that there is a natural generalization of the Jordan-Wigner map in $(2+1)$d compatible with the $\Z_2$ subsystem symmetry. 

On the other hand, it has been noticed in \cite{Seiberg:2020bhn} that the $G$ subsystem symmetry in $(2+1)$d shares many similar features with the $G$ ordinary 0-form symmetry in $(1+1)$d. For example when $G=U(1)$, in both situations, gauging a U(1) ordinary (subsystem) symmetry of a (1+1)D ((2+1)D) QFT results in a U(1) quantum ordinary (subsystem) symmetry in the gauged theory. Moreover, in the context of the ordinary (plaquette) XY model, there exists a mixed anomaly between the ordinary (subsystem) U(1) momentum and winding symmetries. Similar parallel features also exist when $G=\Z_2$. 

Motivated by this observation, we explore whether there exists a $(2+1)$d $\Z_2$ subsystem-symmetry counterpart of the well-known exact boson-fermion duality in $(1+1)$d with a $\Z_2$ ordinary symmetry. The answer turns out to be affirmative. In particular, we propose a family of generalized Jordan-Wigner transformations and apply them to arbitrary lattice models with (on-site) $\Z_2$ subsystem symmetry. The resulting models are fermionic theories with $\Z_2^F$ subsystem fermion parity symmetry. Our generalized Jordan-Wigner transformation was already discussed in \cite{Tantivasadakarn:2020lhq}, but only the mapping of local operators was discussed there. In the present work, we focus on the mapping of the global twist sectors and symmetry sectors, and thus establish the exact boson-fermion duality with global sectors properly matched. We also discuss couplings to background gauge fields and identify the topological terms. As a by-product, we propose a novel $(2+1)$d subsystem generalization of the conventional Arf invariant in $(1+1)$d, and study its foliation structure.   

\paragraph{Organization of the paper:}
In Section \ref{sec.JW2d}, we revisit the boson-fermion duality with an ordinary $\Z_2$ symmetry in $(1+1)$d. We present the discussion in a way that is generalizable to cases with $\Z_2$ subsystem symmetry in $(2+1)$d. In Section \ref{sec:3d}, we proceed to the study the exact boson-fermion duality with $\Z_2$ subsystem symmetry in $(2+1)$d. In Section \ref{sec:3darf}, we discuss the property, in particular the foliation structure, of the subsystem Arf invariant.  In Section \ref{sec:app3d}, we apply the general result to concrete lattice models. 
Moreover, although the resulting quartic fermion theory is not exactly solvable,
we can show that the partition function of the theory vanishes exactly. We conclude this paper in Section \ref{sec:comments} with future directions. 

\section{\texorpdfstring{Boson-fermion duality with $\Z_2$ symmetry in $(1+1)$d revisited}{Boson-fermion duality with Z2 symmetry in (1+1)d revisited}}
\label{sec.JW2d}

In this section, we review the boson-fermion duality in $(1+1)$d. The duality between the $\Z_2$ gauged critical Ising spin chain and the Kitaev Majorana fermion chain is the most well-known example. We first review the derivation of the \emph{exact} duality of the bosonic and fermionic models on the lattice, where the two dual models are related by the  Jordan-Wigner mapping. Then we take the long-distance limit, and find the boson-fermion duality between continuum field theories \cite{Ji:2019ugf,Karch:2019lnn,Hsieh:2020uwb,Fukusumi:2021zme,Ebisu:2021acm}. 

\subsection{Exact duality from the Jordan-Wigner map}
\label{sec:JWmap}

\paragraph{Bosonic System:}
We work on a closed one-dimensional spin chain of size $L$. Each site supports a local state of spin-$\frac{1}{2}$, $\ket{\sigma}_i$ where $\sigma=\pm 1$ and $i=1, ..., L$. The two states $\ket{+1}_i$ and $\ket{-1}_i$ span a two-dimensional Hilbert space at site $i$. The state $\ket{\sigma}_i$ can be acted upon by the Pauli matrices $X_i, Y_i, Z_i$ in the canonical way:
\begin{align}\label{XZaction}
\begin{split}
    X_i\ket{\sigma}_i=\ket{-\sigma}_i, \hspace{1cm}
    Z_i\ket{\sigma}_i=\sigma\ket{\sigma}_i,
\end{split}
\end{align}
and the action of $Y_i$ is fully determined by $Y_i=i X_i Z_i$. 
The spin-$\frac{1}{2}$'s satisfy the boundary condition
\begin{eqnarray}\label{2dBC}
\ket{\sigma}_{i+L}= \ket{t \sigma}_i, \hspace{1cm} t=\pm 1,
\end{eqnarray}
where $t=1$ represents the periodic boundary condition (PBC), while $t=-1$ represents the anti-periodic boundary condition (ABC). Compatibility between \eqref{XZaction} and \eqref{2dBC} requires the boundary conditions of the Pauli operators, 
\begin{eqnarray}\label{XZbc}
X_{i+L}= X_{i}, \hspace{1cm} Z_{i+L}= t Z_{i}.
\end{eqnarray}
We further demand that the system has a $\Z_2$ global symmetry, whose generator is given by
\begin{eqnarray}
U=\prod_{i=1}^{L} X_i.
\end{eqnarray}
We will denote the eigenvalue of the $\Z_2$ generator $U$ by $u$. The signs $(t,u)$ label the twist and \twine sectors.

Using the symmetry eigenvalue $u=\pm 1$ and the boundary condition $t=\pm 1$, the Hilbert space  of any $\Z_2$ symmetric system splits into four twist and \twine  sectors which we label as $\mS,\mT,\mU,\mV$ respectively: $\mS\leftrightarrow (u,t)=(1,1)$, $\mT \leftrightarrow (u,t)=(-1,1)$, $\mU \leftrightarrow (u,t)=(1,-1)$, $\mV \leftrightarrow (u,t)=(-1,-1)$ \cite{Hsieh:2020uwb}. The corresponding partition function in each sector is obtained by inserting suitable $\Z_2$ topological defects in the spatial or temporal cycle:
\begin{align}
\begin{split}
    Z_{\text{bos}}[u=1,t=1]&=Z_{\mS}= \Tr_{\CH_{\text{PBC}}} \frac{1+U}{2}e^{-\beta H_{\text{bos}}},\\
    Z_{\text{bos}}[u=-1,t=1]&=Z_{\mT} = \Tr_{\CH_{\text{PBC}}} \frac{1-U}{2}e^{-\beta H_{\text{bos}}},\\
    Z_{\text{bos}}[u=1,t=-1]&=Z_{\mU} = \Tr_{\CH_{\text{ABC}}} \frac{1+U}{2}e^{-\beta H_{\text{bos}}},\\
    Z_{\text{bos}}[u=-1,t=-1]&=Z_{\mV} = \Tr_{\CH_{\text{ABC}}} \frac{1-U}{2}e^{-\beta H_{\text{bos}}}.
\end{split}
\end{align}

\paragraph{Fermionic System:}
We further consider an arbitrary fermion system with a $\Z_2^F$ fermion parity symmetry. We again assume the system is defined on a closed chain with $L$ sites. Each site supports a local two-dimensional Hilbert space spanned by $\ket{n}_i$, where $n=0,1$ is the fermion number. A complex fermion operator $c_i$ can act on $\ket{n}_i$ in the standard way: $c_i\ket{0}_i=0, \ket{1}_i= c_i^\dagger \ket{0}_i$ and $c_i^\dagger \ket{1}_i=0$. The fermion number operator is $n_i= c_i^\dagger c_i$, and for simplicity we also denote its eigenvalue by the same symbol. Any fermionic system has a $\Z_2^F$ fermion parity symmetry, generated by 
\begin{eqnarray}
P_f=\exp\left({i \pi \sum_{k=1}^{L}n_k}\right),
\end{eqnarray}
and we define its eigenvalue as $u_f$. It is also useful to introduce real fermions
\begin{eqnarray}
\gamma_j= c_j+ c_j^\dagger, ~~~~~\gamma'_j= (c_j-c_j^\dagger)/i,
\end{eqnarray}
which satisfy the standard anti-commutation algebra $\{\gamma_i, \gamma_j\}= 2\delta_{ij}$, $\{\gamma'_i, \gamma'_j\}= 2\delta_{ij}$, and $\{\gamma_i, \gamma'_j\}= 0$. The fermion number operator then becomes $n_k=\frac{1}{2}(1+i \gamma_k \gamma'_k)$. The boundary condition of the state can be either NS (anti-periodic) or R (periodic),
\begin{align}
\ket{n}_{i+L} = 
\begin{cases}
(-1)^{n} \ket{n}_{i}, & \hspace{0.5cm} \text{NS},\hspace{0.3cm} t_f=1,\\
\ket{n}_i, & \hspace{0.5cm} \text{R},\hspace{0.3cm} t_f=-1.
\end{cases} 
\end{align}
This also induces boundary conditions on the fermionic operators
\begin{align}\label{fermionbc}
\begin{split}
    &c_{i+L}=-t_f c_{i}, \hspace{1cm} c_{i+L}^\dagger=-t_f c_{i}^\dagger,\\
    &\gamma_{i+L}= -t_f \gamma_i, \hspace{1cm} \gamma_{i+L}' = -t_f \gamma_{i}'.
\end{split}
\end{align}

As in the case of the bosonic system, the Hilbert space of the fermionic system can also be divided into four twist and \twine sectors labeled by $(u_f, t_f)\in (\pm 1, \pm 1)$. The corresponding partition function in each sector is obtained by inserting suitable $\Z_2^{F}$ fermion parity topological defects in the spatial or temporal cycle:
\begin{align}
\begin{split}
    Z_{\text{fer}}[u_f=1,t_f=1]&= \Tr_{\CH_{\text{NS}}} \frac{1+P_f}{2}e^{-\beta H_{\text{fer}}},\\
    Z_{\text{fer}}[u_f=-1,t_f=1]&= \Tr_{\CH_{\text{NS}}} \frac{1-P_f}{2}e^{-\beta H_{\text{fer}}},\\
    Z_{\text{fer}}[u_f=1,t_f=-1]&= \Tr_{\CH_{\text{R}}} \frac{1+P_f}{2}e^{-\beta H_{\text{fer}}},\\
    Z_{\text{fer}}[u_f=-1,t_f=-1]&= \Tr_{\CH_{\text{R}}} \frac{1-P_f}{2}e^{-\beta H_{\text{fer}}}.
\end{split}
\end{align}
We introduce below a JW map that relates these four sectors with the sectors in the bosonic system. 

\begin{table}[t]
    \centering
    \begin{minipage}{.5\linewidth}
      \centering
        \begin{tabular}{|c|cc|}
        \hline
             \text{boson}  & $t=1$ & $t=-1$ \\
\hline
$u=1$ & \cellcolor{lightblue}  $\mS$  & \cellcolor{lightpink}  $\mU$  \\
$u=-1$ & \cellcolor{lightgreen}  $\mT$  &  \cellcolor{lightyellow}  $\mV$ \\
\hline
        \end{tabular}
    \end{minipage}%
    \begin{minipage}{.5\linewidth}
      \centering
        \begin{tabular}{|c|cc|}
        \hline
            \text{Fermion}  & $t_f=1$ & $t_f=-1$ \\
\hline
$u_f=1$ & \cellcolor{lightblue}  $\mS$  & \cellcolor{lightpink}  $\mU$   \\
$u_f=-1$ & \cellcolor{lightyellow}  $\mV$ &  \cellcolor{lightgreen}  $\mT$ \\
\hline
        \end{tabular}
    \end{minipage} 
    \caption{Symmetry sectors of the boson theory and the fermion theory with a global $\mathbb Z_2$ symmetry}
    \label{tab:JW2d}
\end{table}

\paragraph{Jordan-Wigner Map:}
The Jordan-Wigner (JW) map is a non-local transformation between the Pauli operators $\{X_i, Z_i\}$ and real fermion operators:
\begin{equation}\label{2dJW}
\renewcommand{\arraystretch}{2}
\begin{array}{cc}
    X_j= -i \gamma_j \gamma_j', \hspace{0.5cm} Z_j= \exp\left(\frac{i \pi}{2} \sum_{j'=1}^{j-1}(1+i \gamma_{j'}\gamma'_{j'})\right)\gamma_j, \hspace{0.5cm} Y_j= -\exp\left(\frac{i \pi}{2} \sum_{j'=1}^{j-1}(1+i \gamma_{j'}\gamma'_{j'})\right)\gamma_j',\\
    \gamma_j= \left(\prod_{j'=1}^{j-1} X_{j'}\right) Z_j,
     \hspace{1cm} \gamma_j'= -\left(\prod_{j'=1}^{j-1} X_{j'}\right) Y_j.
\end{array}
\end{equation}
First, the JW map dictates  $P_f=U$, which implies $u_f=u$.  
The boundary conditions of the fermion operators also follow from those of the Pauli operators, $\gamma_{j+L}=-t u \gamma_j$ and $\gamma_{j+L}'=-t u \gamma_j'$. To see this, we use the definition \eqref{2dJW}, the boundary conditions of Pauli operators \eqref{XZbc} as well as eigenvalue of the symmetry operator $u$, 
\begin{eqnarray}\label{gammader}
\gamma_{j+L}= \left(\prod_{j'=1}^{j+L-1} X_{j'}\right) Z_{j+L} = \left(\prod_{j'=1}^{L} X_{j'}\right) \left( \prod_{j'=L+1}^{j+L-1} X_{j'}\right) t Z_{j} = t \left( \prod_{j'=1}^{L} X_{j'}\right) \gamma_j= t P_f \gamma_j.
\end{eqnarray}
To replace the operator $P_f$ by its eigenvalue, we need to move $P_f$ to the most right, yielding $\gamma_{j+L}= -t \gamma_j P_f = -t u \gamma_j$. 
Similar relation also holds for $\gamma_j'$.  Comparing with \eqref{fermionbc}, we find the relations between sectors in the bosonic and fermionic theories as follows:
\begin{eqnarray}\label{sectormap}
u_f=u, \hspace{1cm} t_f= tu.
\end{eqnarray}
This implies that the sectors $\mS,\mT,\mU,\mV$ in the bosonic theories are mapped to $\mS,\mV,\mU,\mT$ in the fermionic theories. See Table \ref{tab:JW2d}. In terms of the partition function, we have an exact equivalence
\begin{eqnarray}\label{2dJWbosfer}
Z_{\text{bos}}[u,t] \equiv Z_{\text{fer}}[u_f,t_f] \stackrel{\eqref{sectormap}}{=} Z_{\text{fer}}[u,tu],
\end{eqnarray}
where the first equality comes from the JW map \eqref{2dJW} which is an exact rewriting of the same theory in terms of different (and highly mutually non-local) degrees of freedom, and the second equality follows from \eqref{sectormap}. This is the boson-fermion duality written in terms of the symmetry and twist sectors. We would like to emphasize that \eqref{2dJWbosfer} is an exact duality and holds for arbitrary $\Z_2$ symmetric theories. 

\subsection{Coupling to background fields}
\label{sec.arfJW}

The boson-fermion duality we derived using the JW map in Section \ref{sec:JWmap} was expressed in terms of the twist and \twine sectors labeled by ${u,t}\leftrightarrow {u_f, t_f}$. These sectors are also in one-to-one correspondence with the presence/absence of the symmetry defects along each direction.  It is well-known that turning on symmetry defects amounts to activating the background gauge fields for the symmetry.  In this subsection, we work out the explicit gauge field dependence of the partition function and recast the duality \eqref{2dJWbosfer} in terms of the gauge fields. This reproduces the more well-known formulation of the boson-fermion duality in \cite{Karch:2019lnn,Ji:2019ugf}. 

We will be interested in the path-integral formalism in this subsection and assume the spacetime to be a torus that contains $T$ sites along the time direction and $L$ sites along the spatial direction. 
Let us denote the background gauge field for the $\Z_2$ symmetry as $B^t_{i,j}, B^x_{i,j}$, normalized such that $B^t_{i,j}\sim B^t_{i,j}+2, B^x_{i,j}\sim B^x_{i,j}+2$. The subscripts $i,j$ are the spacetime coordinates, with $i\in \Z_T, j\in \Z_L$.   We also denote their holonomies as
\begin{eqnarray}\label{Holo}
W_t^B= \sum_{i=1}^{T} B^{t}_{i,j}, \hspace{1cm} W_x^B= \sum_{j=1}^L B^x_{i,j},
\end{eqnarray}
where each equality is defined modulo 2. Since the background fields $B^{t,x}$ are activated by turning on the symmetry and twist operators respectively, the fact that these operators form closed lines implies that the background fields are flat, i.e. $\Delta_t B^x - \Delta_x B^t=0$.

Let us establish the relation between $\{t,u\}$ and $\{B^x, B^t\}$. First, $t=1,-1$ means PBC and ABC when the spin travels around the spatial cycle respectively, which correspond to $W_x^B=0,1$. Thus we have 
\begin{eqnarray}
t = (-1)^{W_{x}^B}.
\end{eqnarray}
On the other hand,  $(-1)^{W_{t}^B}$, as the holonomy along the time direction, probes whether or not the $\Z_2$ symmetry operator $U$ is inserted along the spatial direction. Let us denote by $Z_{\text{bos}}(B^t,B^x)$ the partition function as a function of the background gauge fields. This partition function should be distinguished from  $Z_{\text{bos}}[u,t]$ as a function of the twist and \twine sectors $u,t$. The two
are related by the relation
\begin{eqnarray}\label{IsingFer2}
Z_{\text{bos}}\left(B^t, B^x\right) = \sum_{u=\pm 1} u^{W_t^B} Z_{\text{bos}}\left[u,(-1)^{W_x^B}\right],
\end{eqnarray}
and the inverse relation
\begin{eqnarray}\label{IsingFer}
Z_{\text{bos}}[u,t]= \frac{1}{2}\sum_{W_t^B=0, 1} u^{W_t^B} Z_{\text{bos}}\left(B^t, B^x\right), \hspace{1cm} W_x^B=\frac{1-t}{2},
\end{eqnarray}
where the last equality is defined modulo 2. We note that the right hand side of \eqref{IsingFer2} depends only on the holonomies of the background field, while on the left hand side, we still demand the partition function to depend on $B^{t,x}$ rather than their holonomies. This is to emphasize that the matter fields (the real scalars) couple directly to $B^{t,x}$ rather than to their holonomies. See Section \ref{sec:continfield2d} for further discussion on this point. 

Let us comment on the analogous relations for the fermionic theories.  We need to replace the $\Z_2$ symmetry and its background fields $B^t, B^x$ by the $\Z_2^F$ fermion parity symmetry and its background fields $A^t, A^x$, and also replace $[u,t]$ by $[u_f, t_f]$.  There is a subtlety. Any fermionic theory depends on the choice of the spin structure $\rho_t$ and $ \rho_x$, which are the boundary conditions of the fermions along the time and spatial cycle respectively. The choice $\rho_{t,x}=0,1$ represents the NS, R boundary condition.  Once the background fields $A^t, A^x$ are turned on, the spin structures are shifted by 
\begin{eqnarray}\label{219}
\rho_t \to \rho_t+ W_t^A, \hspace{1cm} \rho_x\to \rho_x + W_x^A,
\end{eqnarray}
which produce new spin structures. Hence for convenience, we can set both $\rho_{t}$ and $\rho_x$ to be zero (i.e. the NS boundary condition), and convert the spin structure dependence to the $\Z_2^F$ background field dependence, and denote the partition functions as $Z_{\text{fer}}(A^t, A^x)$ and $Z_{\text{fer}}[u_f, t_f]$. They are related by 
\begin{eqnarray}\label{fer1}
Z_{\text{fer}}\left(A^t, A^x\right) = \sum_{u_f=\pm 1} u_f^{W_t^A} Z_{\text{fer}}\left[u_f,(-1)^{W_x^A}\right],
\end{eqnarray}
and the inverse relation
\begin{eqnarray}\label{fer2}
Z_{\text{fer}}[u_f,t_f]= \frac{1}{2}\sum_{W_t^A=0, 1} u_f^{W_t^A} Z_{\text{fer}}\left(A^t, A^x\right), \hspace{1cm} W_x^A=\frac{1-t_f}{2}.
\end{eqnarray}

Having established how the bosonic and fermionic theories depend on the background fields, we proceed to convert the boson-fermion duality \eqref{2dJWbosfer} in terms of $\{u,t,u_f,t_f\}$ to a more familiar duality in terms of the gauge fields. 
Combining \eqref{IsingFer}, \eqref{IsingFer2}, \eqref{fer1}, \eqref{fer2} and \eqref{2dJWbosfer}, we find 
\begin{align}\label{IsingFer4}
\begin{split}
    Z_{\text{fer}}(A^t, A^x) &= \sum_{u_f=\pm 1} u_f^{W_t^A} Z_{\text{fer}}\left[u_f, (-1)^{W_x^A}\right]\\
    &= \sum_{u_f=\pm 1} u_f^{W_t^A} Z_{\text{bos}} \left[u_f, (-1)^{W_x^A} u_f\right]\\
    &= \frac{1}{2}\sum_{u_f=\pm 1} u_f^{W_t^A} \sum_{W_t^b=0,1} u_f^{W_{t}^b} Z_{\text{bos}}\left( b^t, b^x\right)\bigg|_{W_x^b= \frac{1-(-1)^{W_x^A}u_f}{2}}\\
    &= \frac{1}{2}\sum_{W_t^b,W_x^b=0,1}(-1)^{(W_x^A+W_x^b)(W_t^A+W_t^b)}Z_{\text{bos}}(b^t,b^x),
\end{split}
\end{align}
where in the third line we have introduced $b^x$ which satisfies $(-1)^{W_x^b}= (-1)^{W_x^A} u_f$,
and in the last line we replaced the sum over $u_f=\pm 1$ by the sum over $W_x^b=0,1$ using the relation
$u_f = (-1)^{W_x^A+W_x^b}$.  Note that in the last step,  the coefficient within the sum is nothing but the Arf invariant of the gauge field $A+b$, 
\begin{eqnarray}\label{2darf}
(-1)^{\text{Arf}(A+b)} :=  (-1)^{(W_x^A+W_x^b)(W_t^A+W_t^b)}.
\end{eqnarray}
We emphasize that the Arf invariant is not an integral of a local Lagrangian density. Instead it is highly non-local, which is expressed in terms of the holonomies of the gauge fields. Moreover, $\text{Arf}(A+b)=1$ only when the holonomies of $A+b$ along both directions are non-trivial, i.e. the RR spin structure. Otherwise, $\text{Arf}(A+b)=0$. This is consistent with the standard definition of the Arf invariant. In summary, comparing the first and last line of  \eqref{IsingFer4}, we find the boson-fermion duality in terms of the gauge fields to be
\begin{eqnarray}\label{bosfer2d}
Z_{\text{fer}}(A^t,A^x) = \frac{1}{2}\sum_{W_t^b,W_x^b=0,1} (-1)^{\Arf(A+b)}Z_{\text{bos}}(b^t,b^x),
\end{eqnarray}
which is precisely of the same duality
discussed in detail in \cite{Karch:2019lnn, Ji:2019ugf, Hsieh:2020uwb, Fukusumi:2021zme, Yao:2020dqx,Yao:2019bub,Ebisu:2021acm}. 
Let us make several comments:
\begin{enumerate}
    \item The Arf invariant is the topological action of the non-trivial phase of the Kitaev Majorana chain. Hence we also use $\Kitaev$ to schematically represent the Arf invariant. 
    \item From \eqref{bosfer2d}, it is transparent that the fermion theory is dual to the \emph{fermionization} of the boson theory. Here, the fermionization is defined by first stacking an Arf invariant onto the bosonic theory and then gauging the diagonal $\Z_2$ symmetry of the theory. Schematically, \eqref{bosfer2d} is represented as 
    \begin{eqnarray}\label{2dferboskitaev}
    \text{Fer} = \text{fermionization}(\text{Bos}):= \frac{\text{Bos}\otimes \Kitaev}{\Z_2}.
    \end{eqnarray}
    Hence the global symmetries match across the duality. In particular, both sides of the duality depend on the spin structure (labeled by $A^{t}, A^{x}$ as discussed below \eqref{219}): the spin structure on the left side comes from the boundary condition of the fermions, and that on the right side comes from $\Kitaev$. 
    \item In terms of the twist and \twine sectors, the fermionization operation is reflected by exchanging the sectors $\mT\leftrightarrow \mV$ as shown in Table \ref{tab:JW2d}. 
    \item The boson-fermion duality is exact, since the duality \eqref{bosfer2d} and \eqref{2dJWbosfer} is obtained by applying the exact JW map. However, in the next subsection, we will consider a boson-fermion duality between continuum QFTs, and it should be treated as an infrared duality. 
    \item The duality can be applied to an arbitrary spin lattice model, even without a $\Z_2$ symmetry. However, in general, only when the system has a non-anomalous $\Z_2$ global symmetry (where the symmetry generator can be represented by an on-site operator $\prod X_i$) can the locality be preserved. For example, if $\Z_2$ is not a symmetry, then one can not replace $\prod_i X_i$ by its eigenvalue $u$ in \eqref{gammader}. This point will become more transparent when discussing the $\Z_2$ subsystem-symmetric systems in Section \ref{sec:3d}. 
    \item The boson-fermion duality \eqref{bosfer2d} can be naturally generalized to a general oriented 2d spacetime of genus $g$. One simply replaces the overall coefficient by $1/2^g$, and the summation by $b\in H^1(M_2, \Z_2)$. The definition of the Arf invariant on general manifolds was defined in \cite{ASENS_1971_4_4_1_47_0} (see also Eq.(II.1) in \cite{Ji:2019ugf}.)
\end{enumerate}

\subsection{Application to the Ising-Majorana duality}
\label{sec:continfield2d}

Having established the boson-fermion duality in general, let us apply it to a concrete lattice model. We will revisit the duality between the Ising model and the Majorana fermions. The discussion here is well-known, but we revisit it as a warm-up exercise for more complicated models in Section \ref{sec:3d}. 

\subsubsection{The exact duality on the lattice}

The Hamiltonian of the critical Ising model is
\begin{eqnarray}\label{IsingHam}
H_{\text{Ising}}=-\sum_{i=1}^{L-1} Z_{i} Z_{i+1} -\sum_{i=1}^{L} X_{i} - t Z_{L}Z_{1},
\end{eqnarray}
where we used the boundary condition \eqref{XZbc} for the last term $-  Z_{L}Z_{L+1}=- t Z_{L}Z_{1}$. Substituting  the JW map \eqref{2dJW} into the Ising model Hamiltonian \eqref{IsingHam}, we obtain the Hamiltonian of the free Majorana fermions
\begin{eqnarray}\label{FermiHam}
H_{\text{Maj}}=\sum_{j=1}^{L-1} i\gamma_{j}'\gamma_{j+1} +\sum_{j=1}^{L} i\gamma_{j}\gamma_{j}' - t u i\gamma_{L}'\gamma_{1}.
\end{eqnarray}
This implies that $t_f=t u$,  as expected in \eqref{sectormap} and Table \ref{tab:JW2d}. (Note that $t_f=1$ is the NS (anti-periodic) boundary condition.)   In the convention of \cite{Hsieh:2020uwb, Fukusumi:2021zme,Ebisu:2021acm}, this is the fermionization map
\begin{eqnarray}\label{fermionization}
\text{Majorana} = \frac{\text{Ising}\times \Kitaev}{\Z_2}.
\end{eqnarray}

\subsubsection{The infrared duality in the continuum}
\label{sec:2dcont}

The duality between lattice models can be realized also in the continuum limit between the proper continuum field theories, as has been discussed in \cite{Karch:2019lnn,Ji:2019ugf,Hsieh:2020uwb, Fukusumi:2021zme,Ebisu:2021acm}. We will sketch the process of taking the continuum limit and discuss the corresponding duality. 

First, it is well-known (see \cite{Karch:2019lnn} for review and references therein) that the continuum limit of the critical Ising model \eqref{IsingHam} under the PBC is given by a Wilson-Fisher fixed point of a single real non-compact scalar field
\begin{eqnarray}\label{Isingfield}
\mathcal{L}_{\text{Ising}}=\frac12(\partial_\mu\phi)^2+\phi^4.
\end{eqnarray}
Here $\phi(x)$ is a real non-compact scalar.  
It is straightforward to see that the kinetic term in the spatial direction comes from the $ZZ$ term in the lattice Ising model,
\begin{eqnarray}
-Z_{i}Z_{i+1}=\frac12(Z_{i+1}-Z_{i})^2-2 \simeq  \frac12(\partial_{x}\phi)^2,
\end{eqnarray}
and $\phi$ is the continuum limit of $Z_i$. We dropped a constant piece which simply shifts the entire energy spectrum. 
The time-derivative component of the kinetic term in the field theory is harder to see from the lattice model. One way to motivate it is by imposing the Lorentz invariance upon taking the continuum limit. The $\phi^4$ term is required to stabilize the theory, but its connection to the lattice model is less clear. We denote the partition function of \eqref{Isingfield} by $\widehat{Z}_{\text{Ising}}(B^t=0,B^x=0)$. Further coupling to the background field is straightforward: we simply replace the ordinary derivative $\partial_{\mu}$ by the covariant derivative $D_{\mu}= \partial_{\mu}-i \pi B_{\mu}$ in \eqref{Isingfield}.  We use the hat to distinguish it from the partition function for the lattice model $Z_{\text{Ising}}(B^t, B^x)$. However, one would expect that they coincide in the long-distance limit. We use $\longleftrightarrow$ to connect the two objects which coincide in the long-distance limit,
\begin{eqnarray}\label{bbduel}
Z_{\text{Ising}}(B^t, B^x) \longleftrightarrow \widehat{Z}_{\text{Ising}}(B^t,B^x).
\end{eqnarray}

We next consider the continuum limit of the  Majorana fermion model \eqref{FermiHam} under the NS-NS boundary condition.  The proper continuum process is
\begin{eqnarray}
i\gamma'_j\gamma_{j+1}+i\gamma_j\gamma'_{j}=i\gamma'_j(\gamma_{j+1}-\gamma_j)\to i\chi_2\partial_{x}\chi_1,
\end{eqnarray}
where $\chi_1(x)$ and $\chi_2(x)$ are the continuum limit of $\gamma_i$ and $\gamma'_{j}$ respectively. Let us organize $\chi_1$ and $\chi_2$ to be a two component Majorana fermion $\chi$, $\chi= (\chi_1, \chi_2)^T/\sqrt{2}$. With Lorentz invariance 
the continuum field Lagrangian is
\begin{eqnarray}\label{Majlag}
\mathcal{L}_{\text{Maj}}=i\chi^T\Gamma^t\slashed \partial \chi,
\end{eqnarray}
where $\Gamma^t=i\sigma^2$, $\Gamma^x=-\sigma^3$, and $\slashed \partial = \Gamma^\mu \partial_\mu $. We denote the partition function of \eqref{Majlag} by $\widehat{Z}_{\text{Maj}}(A_t=0, A_x=0)$, and that of the lattice model (under NS boundary condition) by $Z_{\text{Maj}}(A_t=0, A_x=0)$. Coupling to the background field of $\Z_2^F$ is straightforward, one just replace $\slashed \partial $ by $\slashed D_A:= \Gamma^{\mu}(\partial_\mu - i \pi A_{\mu})$.  One again expects that in the long distance limit, the two partition functions coincide:
\begin{eqnarray}\label{ffdual}
Z_{\text{Maj}}(A^t, A^x) \longleftrightarrow \widehat{Z}_{\text{Maj}}(A^t,A^x).
\end{eqnarray}

We proceed to apply the boson-fermion duality \eqref{bosfer2d} to establish the duality between field theories in the continuum. 
\begin{align}\label{continuumduality1}
\begin{split}
    \widehat{Z}_{\text{fer}}(A^t, A^x)&\longleftrightarrow Z_{\text{fer}}(A^t, A^x)
    =\frac{1}{2}\sum_{W_t^b,W_x^b=0,1} (-1)^{\Arf(A+b)}Z_{\text{bos}}(b^t,b^x)\\
    &\longleftrightarrow \frac{1}{2}\sum_{W_t^b,W_x^b=0,1} (-1)^{\Arf(A+b)}\widehat{Z}_{\text{bos}}(b^t,b^x),
\end{split}
\end{align}
where we used \eqref{ffdual}, \eqref{bosfer2d} and \eqref{bbduel} successively.
In terms of the action, \eqref{continuumduality1} amounts to
\begin{eqnarray}\label{continuumduality2}
\int_{M_2} i\chi^T\Gamma^t\slashed D_{A}\chi \longleftrightarrow  \int_{M_2} \left(\frac12(D_{b}\phi)^2+\phi^4\right) + \pi \Arf(A+b).
\end{eqnarray}
We emphasize that the boson-fermion duality \eqref{continuumduality1} (or \eqref{continuumduality2}) is not an exact duality, but is an infrared duality.  This infrared duality has been extensively discussed in \cite{Karch:2019lnn}, where many nontrivial consistency tests were performed, including matching the nearby gapped phases, matching the anomalies, and matching the operator quantum numbers. Here, we motivate this infrared duality by first proving the exact duality \eqref{bosfer2d} on the lattice, and then taking the continuum limit. 

\subsubsection{Exactly vanishing fermionic partition function}
\label{sec:2dexact}

One of the main advantages of the boson-fermion duality between the critical Ising model and Majorana fermion model is the exact solvability of the theory:
while the critical Ising model does not appear to be exactly solvable, its dual, the Majorana fermion, is a free field theory and is exactly solvable. 

In this section, we show that when the spin structures along the time and spatial directions are both R, the partition function of the Majorana fermion is exactly zero. Although this result trivially follows from the exact solvability of the Majorana fermion, we will generalize the same method to 
higher-dimensional case in section \ref{sec:quarticfermion}, where the model is not exactly solvable. 

Let us start with the Majorana fermion model on the lattice \eqref{FermiHam}. The Hamiltonian is 
\begin{eqnarray}\label{MajHam2d}
H_{\text{Maj}}= \sum_{j=1}^{L-1} i\gamma_{j}'\gamma_{j+1} +\sum_{j=1}^{L} i\gamma_{j}\gamma_{j}' - t_f i\gamma_{L}'\gamma_{1},
\end{eqnarray}
where $t_f=\pm 1$ represents the NS, R spin structure along the spatial direction respectively. The symmetry operator is 
\begin{eqnarray}\label{Majsym2d}
P_f=\prod_{j=1}^{L} \left( - i \gamma_j \gamma'_j\right), 
\end{eqnarray}
whose eigenvalue is $P_f=u_f$. We would like to show that the partition function with the R spin structure on both time and spatial directions vanishes, i.e.\ (recall \eqref{fer1})
\begin{eqnarray}\label{RR}
Z_{\text{Maj}}^{\text{RR}}= Z_{\text{Maj}}[u_f=1, t_f=-1]- Z_{\text{Maj}}[u_f=-1, t_f=-1] =0.
\end{eqnarray}

To see this we consider the following transformation of the fermion operators, 
\begin{align}\label{fftrans2d}
\begin{split}
    \widetilde{\gamma}_j'&= \gamma_{j+1}, \hspace{1cm} j=1, ..., L-1,\\
    \widetilde{\gamma}'_L&= -t_f \gamma_1,\\
    \widetilde{\gamma}_j&= \gamma_j', \hspace{1cm} j=1, ..., L.
\end{split}
\end{align}
Under \eqref{fftrans2d}, the Hamiltonian \eqref{MajHam2d} maps another Hamiltonian of the same form, but with $\gamma$'s replaced by $\widetilde{\gamma}$'s,
\begin{eqnarray}\label{Majham2d2}
\widetilde{H}_{\text{Maj}}=  \sum_{j=1}^{L-1} i\widetilde{\gamma}_{j}'\widetilde{\gamma}_{j+1} +\sum_{j=1}^{L} i\widetilde{\gamma}_{j}\widetilde{\gamma}_{j}' - t_f i\widetilde{\gamma}_{L}'\widetilde{\gamma}_{1}.
\end{eqnarray}
The symmetry operator \eqref{Majsym2d} is mapped to 
\begin{eqnarray}
\widetilde{P}_f= t_f P_f, 
\end{eqnarray}
which means $\widetilde{u}_f= t_f u_f$. Since \eqref{MajHam2d} and \eqref{Majham2d2} are related via a change of variables \eqref{fftrans2d}, their partition functions should be the same: 
\begin{eqnarray}
Z_{\text{Maj}}[u_f, t_f]= Z_{\widetilde{\text{Maj}}}[u_f t_f, t_f]\equiv Z_{\text{Maj}}[u_f t_f, t_f],
\end{eqnarray}
where the second equality follows from the fact that \eqref{MajHam2d} and \eqref{Majham2d2} share exactly the same form, i.e. they are the same theory. When $t_f=1$, i.e. in the NS sector, the above equality is trivially satisfied. More interesting case is when $t_f=-1$, which is equivalent to \eqref{RR}. Hence the partition function of the fermion with the RR spin structure is exactly zero. 

Let us make some comments. 
\begin{enumerate}
    \item The transformation \eqref{fftrans2d} between two sets of fermion operators is actually the  composition $\text{JW}\circ \text{KW}\circ \text{JW}$ acting on a  fermion theory. In the Language of \cite{Hsieh:2020uwb}, \eqref{fftrans2d} maps $\mathsf{F}\xrightarrow{\text{JW}} \mathsf{A}\xrightarrow{\text{KW}} \mathsf{D}\xrightarrow{\text{JW}} \mathsf{F}'$, where KW is short for the Kramers-Wannier duality transformation. 
    \item The above derivation is equivalent to the fact that a massless Majorana fermion is invariant under stacking an Arf invariant, 
    \begin{eqnarray}\label{arfstacking}
    Z_{\text{Maj}}(A^t, A^x)= (-1)^{\Arf(A^t, A^x)}Z_{\text{Maj}}(A^t, A^x).
    \end{eqnarray}
    Hence the partition function coupled to the background field vanishes precisely when the Arf invariant is nontrivial, i.e. the spin structure is RR. 
    \item The relation \eqref{arfstacking} is a manifestation of the mixed anomaly between the $\Z_2^F$ fermion parity symmetry and the $\Z_2^{F_{L}}$ chiral symmetry. 
\end{enumerate}

\section{\texorpdfstring{Boson-fermion duality with $\Z_2$ subsystem symmetry in $(2+1)$d}{Boson-fermion duality with Z2 subsystem symmetry in (2+1)d}}
\label{sec:3d}

In Section \ref{sec.JW2d},  we used the JW map to establish an exact duality between bosonic and fermionic lattice models. We also revisited the infrared boson-fermion duality for the Ising and Majorana fermion systems by taking the continuum limit. 

In this section,  we will consider systems with $\Z_2$ subsystem symmetries in $(2+1)$d, in parallel to our previous discussion for $(1+1)$d systems.  We first propose a generalized JW transformation, and derive an exact duality both in terms of the twist and \twine sectors $[u,t, u_f, t_f]$ and in terms of the gauge fields. We then apply the exact boson-fermion duality to a concrete lattice model, i.e. the plaquette Ising model, and show that it is dual to a fermion model with a $\Z_2^F$ subsystem fermion parity symmetry, which we call the plaquette fermion model. 

\subsection{Exact duality from the generalized JW map}
\label{sec:3dJWmap}

\subsubsection{Bosonic system}

We work on a closed 2d spatial square lattice whose sites are labeled by $(i,j), i=1,2,..., L_x,j=1,2,..., L_y$, with $i\sim i+L_x, j\sim j+L_y$. Each site supports a local state of spin-$\frac{1}{2}$ $\ket{\sigma}_{i,j}$ where $\sigma_{i,j}=\pm 1$. The two states $\ket{\pm 1}_{i,j}$ span a two-dimensional Hilbert space. The Hilbert space can be acted upon by the Pauli matrices $X_{i,j}, Y_{i,j},Z_{i,j}$ in the canonical way 
\begin{eqnarray}\label{XZ3d}
X_{i,j}\ket{\sigma}_{i,j}= \ket{-\sigma}_{i,j}, \hspace{1cm} Z_{i,j}\ket{\sigma}_{i,j}= \sigma\ket{\sigma}_{i,j}.
\end{eqnarray}
The spin-$\frac{1}{2}$'s satisfy the boundary condition 
\begin{eqnarray}\label{Twistbc3d}
\ket{\sigma}_{i+L_x, j}= \ket{t_{j}^x \sigma}_{i,j}, \hspace{1cm} \ket{\sigma}_{i, j+L_y}= \ket{t_{i}^y \sigma}_{i,j}, \hspace{1cm} \ket{\sigma}_{i+L_x, j+L_y} = \ket{t^{xy} t_j^x t_i^y \sigma }_{i,j},
\end{eqnarray}
where $t_j^x, t_i^y, t^{xy}= \pm 1$ label the twisted boundary conditions along the $j$-th row, $i$-th column and at the corner respectively. The twisted boundary conditions labeled by $t^x_j$ and $t^y_i$ are straightforward generalization of those of the 1d spin chain discussed in \eqref{2dBC}. However, the additional twist parameter $t^{xy}$ is new. To motivate this, we first consider $\ket{\sigma}_{i+L_x, j+L_y}= \ket{t^x_{j+L_y} \sigma}_{i, j+L_y}= \ket{t^x_{j+L_y} t_i^{y}\sigma}_{i,j}$. Alternatively, we also have $\ket{\sigma}_{i+L_x, j+L_y}= \ket{t^y_{i+L_x} \sigma}_{i+L_x, j}= \ket{t^y_{i+L_x} t_j^{x}\sigma}_{i,j}$. Assuming $t^x_{j+L_y}= s^x t^x_{j}$ and $t^y_{i+L_x}= s^y t^y_i$, compatibility between the two paths requires $t^x_{j+L_y} t_i^{y}= t^y_{i+L_x} t_j^{x}$, which is equivalent to $s^x=s^y := t^{xy}$. This justifies the last condition in \eqref{Twistbc3d}.
Compatibility between \eqref{XZ3d} and \eqref{Twistbc3d} requires the boundary conditions of the Pauli operators,
\begin{equation}\label{bcXZ3d}
\renewcommand{\arraystretch}{1.5}
\begin{array}{llll}
    &X_{i+L_x,j}= X_{i,j}, \hspace{1.5cm} &X_{i,j+L_y}= X_{i,j}, \hspace{1.5cm} &X_{i+L_x,j+L_y}= X_{i,j},\\
    &Z_{i+L_x,j}= t^x_j Z_{i,j}, \hspace{1.5cm} &Z_{i,j+L_y}= t^y_i Z_{i,j}, \hspace{1.5cm} &Z_{i+L_x,j+L_y}= t^{xy} t^x_j t^y_i Z_{i,j},
\end{array}
\end{equation}
and the boundary conditions of $Y$ coincides with those of $Z$. Hence there are $2^{L_x+L_y+1}$ twist sectors in total. 

\begin{figure}[htbp]
    \centering
    \includegraphics[width=70mm,scale=0.5]{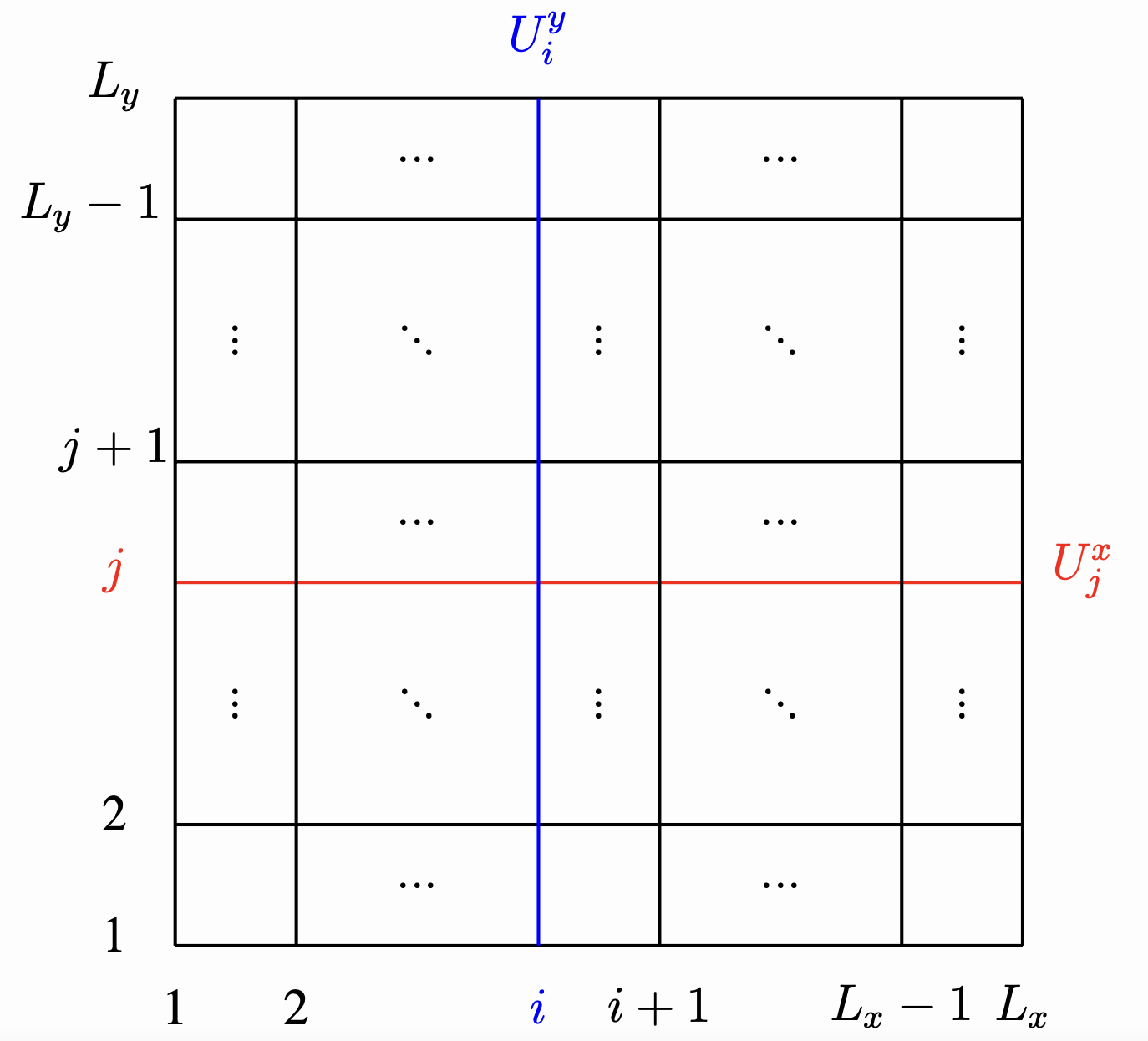}
    \caption{Generators of the $\Z_2$ subsystem symmetry. }
 \label{fig:subgen}
 \end{figure}
 
We further demand that the system has a $\Z_2$ subsystem global symmetry, whose generators are given by 
\begin{align}\label{Z2sub}
\begin{split}
U_j^x=\prod_{i=1}^{L_x}X_{i,j},\hspace{1cm}
U_i^y=\prod_{j=1}^{L_y}X_{i,j}.
\end{split}
\end{align}
See Figure \ref{fig:subgen} for a graphical representation of the symmetry operators.
Let us denote their eigenvalues as $u_j^x, u_i^y$ respectively. Note that these $L_x+L_y$ symmetry operators are not independent, since they satisfy
\begin{eqnarray}\label{trivialU}
\prod_{j=1}^{L_y} U_j^x \prod_{i=1}^{L_x} U_i^y= \prod_{j=1}^{L_y} u_j^x \prod_{i=1}^{L_x} u_i^y=1.
\end{eqnarray}
Hence there are only $L_x+L_y-1$ independent operators, dividing the entire Hilbert space (with a fixed boundary condition) to $2^{L_x+L_y-1}$ sectors. Moreover, similar to the symmetry in $(1+1)$d, the subsystem symmetry is anomaly free since the generators \eqref{Z2sub} are on-site. 

From the above discussion, it appears that the number of twist sectors $2^{L_x+L_y+1}$ is different from the number of \twine sectors $2^{L_x+L_y-1}$. In fact, a more careful discussion shows that there are only $2^{L_x+L_y-1}$ distinguished twist sectors hence the numbers of twist and \twine sectors match. To see this, let us follow the discussion 
below \eqref{FermiHam} and incorporate the twist parameters $\ft=\{t_j^x, t_i^y, t^{xy}\}$ into the Hamiltonian, which we denote as $H_{\text{bos}}[t_j^x, t_i^y, t^{xy}]$. 
Requiring the Hamiltonian to be $\Z_2$ subsystem symmetric yields
\begin{eqnarray}\label{HamU}
H_{\text{bos}}[t_j^x, t_i^y, t^{xy}]= U\cdot  H_{\text{bos}}[t_j^x, t_i^y, t^{xy}]\cdot  U^\dagger .
\end{eqnarray}
On the other hand, from the definition of the twist parameters $\ft$, conjugating the Hamiltonian by $U_j^x$ for any $j=1, ..., L_y$ would flip $t^y_i\to -t^y_i$ for any $i=1, ..., L_x$, and similarly conjugating the Hamiltonian by $U_i^y$ for any $i=1, ..., L_x$ would flip $t^x_j\to -t^x_j$ for any $j=1, ..., L_y$. This means that 
\begin{eqnarray}
H_{\text{bos}}[t_j^x, t_i^y, t^{xy}]= H_{\text{bos}}[-t_j^x, t_i^y, t^{xy}]= H_{\text{bos}}[t_j^x, -t_i^y, t^{xy}] .
\end{eqnarray}
This means that the Hamiltonian depends on $\ft$ only through the quadratic combinations of $\ft$, say $\widehat{t}^x_{j}:=t^x_{j}t^x_{j+1}$ and $\widehat{t}^y_{i}:=t^y_{i}t^y_{i+1}$, which will be discussed in more detail in Section \ref{sec:gaugefieldsbossystems}. In summary this more careful discussion yields $L_x+L_y+1-2=L_x+L_y-1$ independent twist parameters, hence there are $2^{L_x+L_y-1}$ distinct twist sectors, which matches the number of distinct \twine sectors. 

We denote the partition function with the twist sectors labeled by $\mathfrak{t}:= \{t_i^y, t_j^x, t^{xy}\}$ and \twine sectors labeled by $\fu:=\{u_i^y, u_j^x\}$ as  
\begin{eqnarray}\label{Zbosut}
Z_{\text{bos}}[\fu, \ft]= \Tr_{\CH_{\ft}} \left(\prod_{i=1}^{L_x}\frac{1+ u_i^y U_{i}^y}{2} \right)\left(\prod_{j=1}^{L_y}\frac{1+u_j^x U_{j}^x}{2} \right)e^{-\beta H_{\text{bos}}}.
\end{eqnarray}
As we will discuss in the following sections, various choices of $\fu, \ft$ amount to turning on various background gauge fields for the $\Z_2$ subsystem symmetry. 

\subsubsection{Fermionic system} 

We further consider an arbitrary fermion system with a $\Z_2^F$ subsystem fermion parity  symmetry. The spatial lattice is defined in the same way as   in the bosonic case above. Each site supports a local two-dimensional Hilbert space spanned by $\ket{n}_{i,j}$, where $n=0,1$ is the fermion number.  A complex fermion operator $c_{i,j}$ acts on $\ket{n}_{i,j}$ in the standard way: $c_{i,j}\ket{0}_{i,j}=0, \ket{1}_{i,j}=c_{i,j}^\dagger\ket{0}_{i,j}$ and $c_{i,j}^\dagger\ket{1}_{i,j}=0$. The fermion number operator is $n_{i,j}=c_{i,j}^\dagger c_{i,j}$, whose eigenvalue we denote by the same symbol. The generators of the $\Z_2^F$ subsystem fermion parity  symmetry are
\begin{eqnarray}
P_{f, j}^x= \exp\left( i \pi \sum_{i=1}^{L_x} n_{i,j}\right),  \hspace{1cm} P_{f, i}^y= \exp\left( i \pi \sum_{j=1}^{L_y} n_{i,j}\right), 
\end{eqnarray}
and we denote their eigenvalues as $u_{f,j}^x$ and $u_{f,i}^y$ respectively. We also introduce the real fermions for later convenience:
\begin{eqnarray}
\gamma_{i,j}= c_{i,j}+ c_{i,j}^\dagger, \hspace{1cm} \gamma_{i,j}'= (c_{i,j}-c_{i,j}^\dagger)/i.
\end{eqnarray}
The fermion number operator, in terms of the real fermions, is $n_{i,j}= \frac{1}{2}(1+ i \gamma_{i,j}\gamma_{i,j}')$. The boundary conditions are imposed on each row or column, 
\begin{equation}
\begin{split}
\ket{n}_{i+L_x, j}&= 
\begin{cases}
(-1)^{n} \ket{n}_{i,j}, &  \text{NS}, t_{f, j}^x=1,\\
\ket{n}_{i,j},&  \text{R}, t_{f, j}^x=-1,
\end{cases}\\
\ket{n}_{i, j+L_y}&= 
\begin{cases}
(-1)^{n} \ket{n}_{i,j}, &  \text{NS}, t_{f, i}^y=1,\\
\ket{n}_{i,j},&  \text{R}, t_{f, i}^y=-1,
\end{cases}\\
\ket{n}_{i+L_x, j+L_y}&= 
\begin{cases}
\ket{n}_{i,j},&  \text{NS}, t_f^{xy} t_{f, j}^x t_{f,i}^y=1,\\
(-1)^{n} \ket{n}_{i,j}, &  \text{R},t_f^{xy} t_{f, j}^x t_{f,i}^y=-1.
\end{cases}
\end{split}
\end{equation}
These induce the boundary conditions on the fermionic operators,
\begin{eqnarray}\label{3dferbc}
\gamma_{i+L_x,j}= -t_{f,j}^x \gamma_{i,j}, \hspace{1cm} \gamma_{i,j+L_y}= -t_{f,i}^y \gamma_{i,j}, \hspace{1cm} \gamma_{i+L_x, j+L_y} = t^{xy}_{f} t_{f, j}^x t_{f,i}^y \gamma_{i,j}.
\end{eqnarray}
Similar to the discussions below \eqref{Twistbc3d}, we have $t^x_{f,j+L_y}= t^{xy}_{f} t^x_{j}$, and $t^y_{f,i+L_x}= t^{xy}_{f} t^y_{i}$. There are again $2^{L_x+L_y+1}$ twist sectors labeled by $\ft_f:=\{ t^x_{f,j}, t^y_{f,i}, t^{xy}_{f}\}$ (with only $2^{L_x+L_y-1}$ distinguished twist sectors), as well as $2^{L_x+L_y-1}$ \twine sectors labeled by $\fu_f:= \{u^x_{f,j}, u^y_{f,i}\}$ with the constraint $\prod_{j=1}^{L_y} u_{f,j}^x \prod_{i=1}^{L_x} u_{f,i}^y=1$. The partition function labeled by twist  and \twine sectors $[\ft_f,\fu_f]$ is
\begin{eqnarray}
Z_{\text{fer}}[\fu_f, \ft_f]= \Tr_{\CH_{\ft_f}} \left(\prod_{i=1}^{L_x}\frac{1+u_{f,i}^y P_{f,i}^y}{2} \right)\left(\prod_{j=1}^{L_y}\frac{1+u_{f,j}^x P_{f,j}^x}{2} \right)e^{-\beta H_{\text{fer}}}.
\end{eqnarray}

\subsubsection{Generalized JW map}

\begin{figure}[htbp]
    \centering
    \includegraphics[width=60mm,scale=0.5]{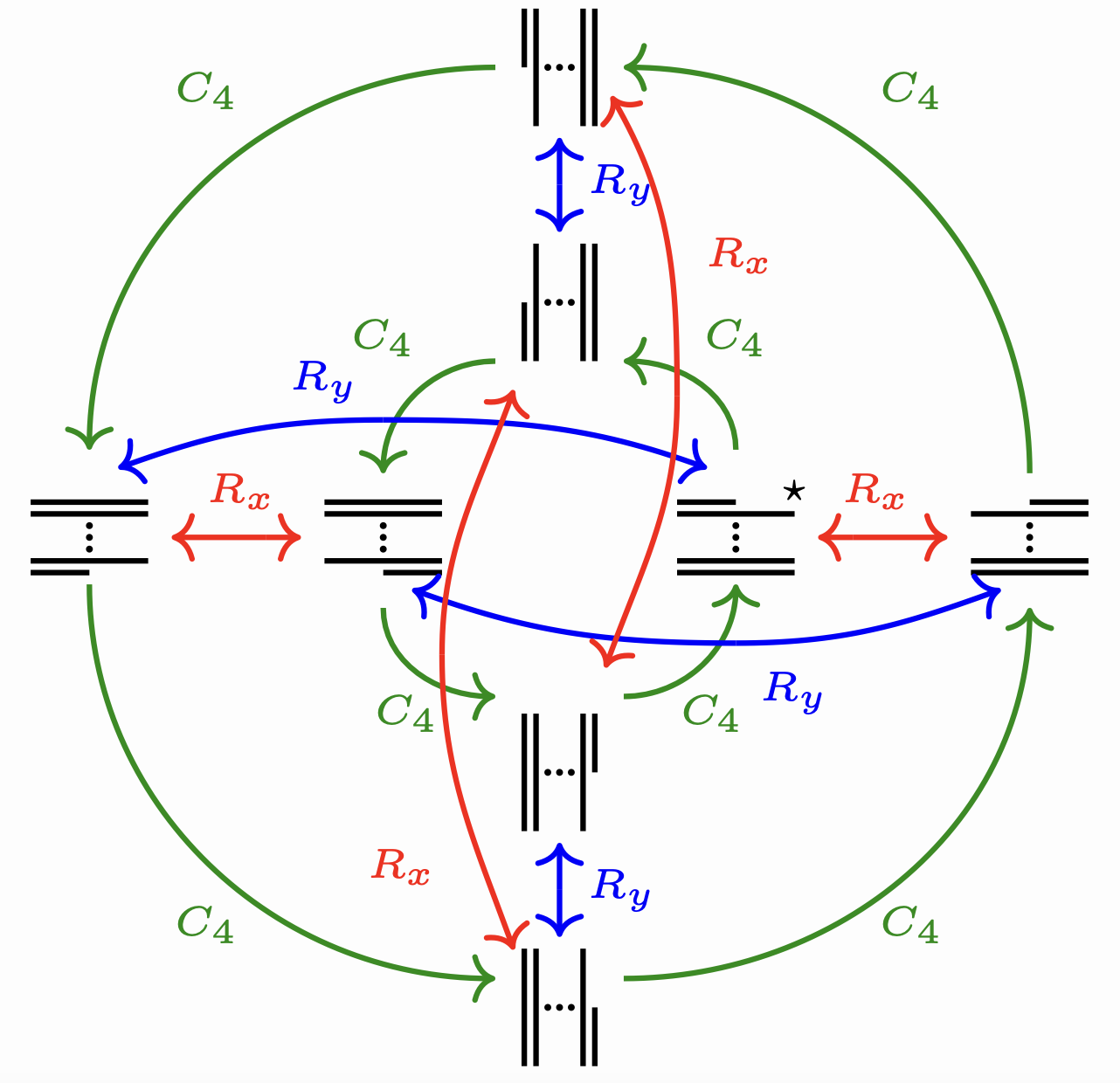}
    \caption{Eight types of JW maps in $(2+1)$d. We only discuss the map marked by the star in detail. }
   \label{fig:8JW}
\end{figure}

It is nontrivial to generalize the JW map in $(1+1)$d to higher dimensions. This is because, in higher dimensions, one needs to specify how to arrange the tail dressed on the fermions to ensure the anti-commuting relation of the fermions. In $(1+1)$d, there are only two choices of dressing the tail to the left or the right of the fermionic operator. The two choices are related by the spatial reflection. However, in $(2+1)$d, we find eight choices of dressing the tails for the fermions, which are related by the  crystallographic space group of the rectangular, i.e.\ the dihedral group $D_4$, 
\begin{eqnarray}
D_4= \{C_4, R_x \, | \, C_4^4=1, R_x^2=1, C_4 R_x= R_x C_4^{-1}\},
\end{eqnarray}
where $C_4$ is the $\pi/2$ rotation around the center, and $R_x$ is the reflection in the $x$ direction. The reflection $R_y$ in the $y$ direction is the composition $R_y=R_x C_4^2$. See Figure \ref{fig:8JW} for a schematic representation of the eight variations of the JW maps. Below, we discuss in detail only one of them, marked by the star in Figure \ref{fig:8JW}. 

\begin{figure}[htbp]
    \centering
    \includegraphics[width=70mm,scale=0.5]{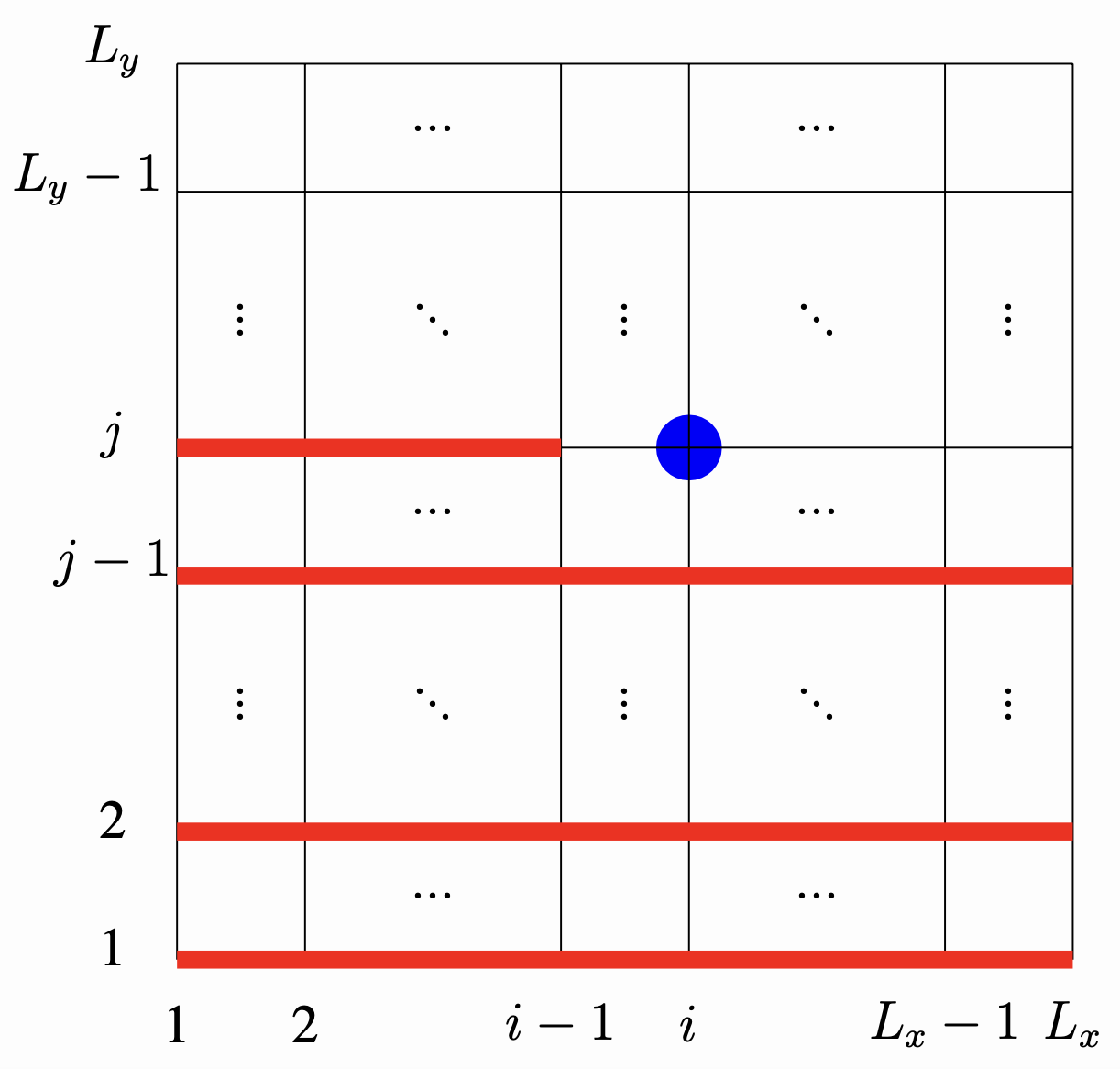}
   \caption{JW maps in $(2+1)$d with the tail winding around $x$ direction. The red line denotes the tail of the product of Pauli $X$ operators and the blue dot denotes the Pauli $Z$ or $Y$ operator.}
    \label{fig:JWx}
\end{figure}

The generalized JW map is defined to be
\begin{align}\label{JWxmaj}
\begin{split}
X_{i,j}&= -i\gamma_{i,j}\gamma'_{i,j},\\
Z_{i,j}&= 
\exp\left[i \pi \left(\sum_{i'=1}^{L_x}\sum_{j'=1}^{j-1}\frac{1+i\gamma_{i',j'}\gamma'_{i',j'}}{2}+\sum_{i'=1}^{i-1}\frac{1+i\gamma_{i',j}\gamma'_{i',j}}{2}\right)\right] \gamma_{i,j},\\
Y_{i,j}&= -
\exp\left[i \pi \left(\sum_{i'=1}^{L_x}\sum_{j'=1}^{j-1}\frac{1+i\gamma_{i',j'}\gamma'_{i',j'}}{2}+\sum_{i'=1}^{i-1}\frac{1+i\gamma_{i',j}\gamma'_{i',j}}{2}\right)\right] \gamma'_{i,j}.
\end{split}
\end{align}
The exponential tail starts from $(1,1)$, winds around the $x$ direction, and stops at $(i-1,j)$, which ensures the commutation relations of the Pauli matrices on different sites. See Figure \ref{fig:JWx} for a pictorial representation. 
The inverse transformation is 
\begin{align}\label{JWxIsing}
\begin{split}
\gamma_{i,j}&=\left(\prod_{i'=1}^{L_x}\prod_{j'=1}^{j-1}X_{i',j'}\right)\left(\prod_{i'=1}^{i-1}X_{i',j}\right)Z_{i,j},\\
\gamma_{i,j}'&=-\left(\prod_{i'=1}^{L_x}\prod_{j'=1}^{j-1}X_{i',j'}\right)\left(\prod_{i'=1}^{i-1}X_{i',j}\right)Y_{i,j},
\end{split}
\end{align}
from which we can also check that the fermion anti-commuting relation is also respected due to the tails. 
The generalized JW map has been discussed in \cite{Tantivasadakarn:2020lhq} where the mapping between local operators are determined. In the bosonic theory, the set of local operators that preserve the $\mathbb Z_2$ subsystem symmetry is generated by $Z_{i,j}Z_{i+1,j}Z_{i,j+1}Z_{i+1,j+1}$ and $X_{i,j}$. This set, via the generalized JW transformation \eqref{JWxIsing}, is mapped to the set of local operators in fermionic theory that preserve the $\mathbb Z_2$ subsystem parity symmetry, generated by $\gamma_{i,j}'\gamma_{i+1,j}\gamma_{i,j+1}'\gamma_{i+1,j+1}$ and $i\gamma_{i,j}\gamma_{i,j}'$. However, a careful treatment of the boundary conditions was not discussed. We will fill this gap below.

Let us proceed to determine how the twist and \twine sectors of the bosonic theories are related to those of the fermionic theories via the generalized JW map \eqref{JWxmaj} and \eqref{JWxIsing}. First, it is straightforward to see the relation between symmetry operators: $U_j^x= P_{f,j}^x$ and $U_i^y=P_{f,i}^y$, which implies $u_{f,j}^x=u_{j}^x, u_{f,i}^y=u_{i}^y$. Furthermore, assuming the boundary condition of the Pauli operators as \eqref{bcXZ3d}, the JW maps induce boundary conditions for the fermions
\begin{equation}
\renewcommand{\arraystretch}{2}
\begin{array}{lll}
    &\gamma_{i+L_x, j}= -u_{j}^x t_{j}^x  \gamma_{i,j}, 
    & \gamma'_{i+L_x, j}= -u_{j}^x t_{j}^x  \gamma'_{i,j}, \\
    &\gamma_{i,j+L_y}= -\left( \prod_{j=1}^{L_y}u_{j}^x\right) t_{i}^y \gamma_{i,j},
    & \gamma'_{i,j+L_y}= -\left( \prod_{j=1}^{L_y}u_{j}^x\right) t_{i}^y \gamma'_{i,j},\\
    &\gamma_{i+L_x,j+L_y}= \left( \prod_{j=1}^{L_y}u_{j}^x\right) u_{j}^x t^{xy} t_{i}^y t_{j}^x \gamma_{i,j}, 
    & \gamma'_{i+L_x,j+L_y}= \left( \prod_{j=1}^{L_y}u_{j}^x\right) u_{j}^x t^{xy} t_{i}^y t_{j}^x \gamma'_{i,j}.
\end{array}
\end{equation}
Note that the factor $\prod_{j=1}^{L_y}u_{j}^x$ is the eigenvalue of the generator of the ordinary $\Z_2$ global symmetry.   
Comparing with \eqref{3dferbc}, we find the relations between the twist and \twine sectors in the bosonic and fermionic theories as follows: 
\begin{equation}\label{3dduality1}
\renewcommand{\arraystretch}{2}
\begin{array}{cc}
u_{f,j}^x=u_{j}^x, \hspace{1cm} u_{f,i}^y=u_{i}^y,\\
t_{f,j}^x= u_j^x t_j^x, \hspace{1cm} t_{f,i}^y=\left( \prod_{j=1}^{L_y}u_{j}^x\right) t_{i}^y, \hspace{1cm} t_f^{xy}= t^{xy}.
\end{array}
\end{equation}
This implies that the sectors of the bosonic theory are permuted in the fermionic theory under the JW map. This is similar to the $(1+1)$d case, although we now have more sectors and the permutation pattern is more complicated. In terms of the partition function, we have the exact equivalence
\begin{equation}\label{3dduality2}
\begin{split}
    Z_{\text{bos}}[u_j^x, u_i^y, t_j^x, t_i^y, t^{xy}]\equiv  Z_{\text{fer}}[u_{f,j}^x, u_{f,i}^y, t_{f,j}^x, t_{f,i}^y, t_f^{xy}] 
    \stackrel{\eqref{3dduality1}}{=}Z_{\text{fer}}\left[u_{j}^x, u_{i}^y, u_j^x t_{j}^x, \left( \prod_{j=1}^{L_y}u_{j}^x\right)t_{i}^y, t^{xy}\right].
\end{split}
\end{equation}
Similar to the situation in $(1+1)$d, the boson-fermion duality \eqref{3dduality2} an \emph{exact} duality which is applicable to arbitrary $(2+1)$d systems with a $\Z_2$ subsystem symmetry. 

\subsection{Coupling to background fields}

The boson-fermion duality we derived using the generalized JW map in Section \ref{sec:3dJWmap} was expressed in terms of the twist and \twine sectors $\fu, \ft, \fu_f, \ft_f$ representing the presence/absence of symmetry defects. In this subsection, we recast the exact boson-fermion duality in terms of gauge fields of the subsystem symmetry. 

\subsubsection{Gauge fields for bosonic systems}
\label{sec:gaugefieldsbossystems}

We will be interested in the path-integral formalism in this subsection, hence we consider the spacetime square lattice (rather than just the spatial square lattice). We assume the time direction contains $T$ sites, and the two spatial directions contain $L_x, L_y$ sites respectively. We first discuss how bosonic theories with $\Z_2$ subsystem symmetry couple to the background gauge fields. 

\paragraph{Time component gauge fields:}
First we can introduce the temporal component of the gauge field $B^t_{k,i,j}$, which lives on the time-like link between sites $(k,i,j)$ and $(k+1, i,j)$. It equals 0 if there are an even number of symmetry operators (along either spatial direction) intersecting the two sites in total, and equals 1 if there are odd.   
We also demand the gauge transformation   
\begin{eqnarray}
B^t_{k,i,j}\to B^t_{k,i,j}+ g_{k+1, i,j}- g_{k, i,j},
\end{eqnarray}
which effectively
moves the spatial-like symmetry operators along the time direction. By choosing a gauge,  we can squeeze all the symmetry generators to one particular time slice.  In other words, we can define a gauge invariant holonomy as in \eqref{Holo}
\begin{eqnarray}
W_{t;i,j}^B=\sum_{k=1}^{T} B^t_{k,i,j},
\end{eqnarray}
which is defined modulo 2. Then acting a symmetry operator $\prod_{j=1}^{L_y} (U_j^x)^{\alpha_j} \prod_{i=1}^{L_x} (U_i^y)^{\beta_i}$ on the ground state amounts to activating the background field with $W_{t;i,j}^B = \alpha_j + \beta_i$. This implies that not all $L_xL_y$ elements of $W_{t;i,j}^B$ are independent. Instead, there are only $L_x+L_y-1$ independent elements, associated with the $L_x+L_y-1$ independent symmetry operators. This motivates the decomposition 
\begin{eqnarray}\label{WBtdef}
W_{t;i,j}^B = W^B_{t,x;j} + W^B_{t,y;i},
\end{eqnarray}
which corresponds to inserting defects $(U^x_j)^{W^B_{t,x;j}} (U^y_i)^{W^B_{t,y;i}}$. Moreover, the global shift
\begin{eqnarray}\label{globalshift}
W^B_{t,x;j}\to W^B_{t,x;j}+1, \hspace{1cm} W^B_{t,y;i}\to W^B_{t,y;i}+1, \hspace{1cm} \forall~ i,j,
\end{eqnarray}
leaves $W^B_{t; i,j}$ invariant for all $i,j$, hence is gauge transformation. For consistently, we can verify that the global shift corresponds to additionally inserting $\prod_{i=1}^{L_x} U_i^y \prod_{j=1}^{L_y} U_j^y$ which is a trivial defect due to \eqref{trivialU}. 

\paragraph{Spatial component gauge fields:}
We further introduce the background field along the spatial direction by the minimal coupling. Note that the $Z$ Pauli operator transforms under the $\Z_2$ \emph{global} subsystem symmetry as $Z_{i,j}\to (-1)^{\alpha_j+ \beta_i} Z_{i,j}$. The minimal combination that is invariant under such a global symmetry transformation is $Z_{i,j} Z_{i+1, j} Z_{i+1, j+1} Z_{i,j+1}$. This is the generalization of the  ordinary differential operator in the presence of the subsystem symmetry on the lattice. We promote the parameters $\alpha_j, \beta_i$ to be coordinate dependent, hence $Z_{i,j}\to (-1)^{g_{i,j}}Z_{i,j}$, which makes the above combination non-invariant.\footnote{We introduced $g_{i,j}= \alpha_j(i,j) + \beta_i(i,j)$ for convenience. }  We then introduce a background gauge field $B^{xy}_{k,i,j}$ to restore the invariance. The gauge transformation of  $B^{xy}_{k,i,j}$ is 
\begin{eqnarray}
B^{xy}_{k,i,j} \to B^{xy}_{k,i,j} + g_{k,i+1,j+1} - g_{k, i+1, j} - g_{k, i, j+1} + g_{k, i, j},
\end{eqnarray}
so that the combination $Z_{i,j} Z_{i+1, j} Z_{i+1, j+1} Z_{i,j+1} (-1)^{B^{xy}_{0,i,j}}$ is gauge invariant.\footnote{We assumed, without loss of generality,  that the Hilbert space where the $Z$'s act on is at time $k=0$. } We also introduce the gauge invariant holomonies of $B^{xy}$ as
\begin{eqnarray}
W_{x;j}^B= \sum_{i=1}^{L_x} B^{xy}_{k, i,j}, \hspace{1cm} W_{y;i}^B= \sum_{j=1}^{L_y} B^{xy}_{k, i,j}.
\end{eqnarray}
Since the holonomies only depend on the presence/absence of the defect line along the time direction which specifies the twisted boundary condition which holds for all time, the holonomies are independent of the time $k$.  The holonomies are determined by the labels of boundary conditions, 
\begin{eqnarray}\label{WBt}
(-1)^{W_{x;j}^B} = t_j^x t_{j+1}^x=:\widehat{t}_j^x, \quad j=1,...,L_y, \hspace{1cm} (-1)^{W_{y;i}^B} = t_{i}^y t_{i+1}^y=:\widehat{t}_i^y \quad i=1,...,L_x.
\end{eqnarray}
The holonomies are not all independent. In particular, they satisfy 
\begin{eqnarray}\label{WBtc}
\prod_{j=1}^{L_y} (-1)^{W_{x;j}^B}= \prod_{j=1}^{L_y} \widehat{t}^x_j= t_1^x t_{L_y+1}^x = t^{xy},\hspace{1cm} \prod_{i=1}^{L_x}(-1)^{W_{y;i}^B}=\prod_{i=1}^{L_x} \widehat{t}^y_i= t_1^y t_{L_x+1}^y = t^{xy}.
\end{eqnarray}
Hence among $L_x+L_y$ holonomies $W_{y;i}^B, W_{x;j}^B$,  only $L_x+L_y-1$ of them are independent.  Here, for convenience, we introduced $\widehat{t}^x_j, \widehat{t}^y_i$, and assumed that the partition function only depends on ${t}^x_j, {t}^y_i$ through their combinations $\widehat{t}^x_j, \widehat{t}^y_i$. See also the discussion below \eqref{HamU}.  Indeed, there is a one-to-one correspondence between the holonomies of the gauge fields and the parameters of the twist sectors $\widehat{t}^x_j, \widehat{t}^y_i, {u}^x_j, {u}^y_i$. 

\begin{figure}[htbp]
    \centering
    \includegraphics[width=50mm,scale=0.5]{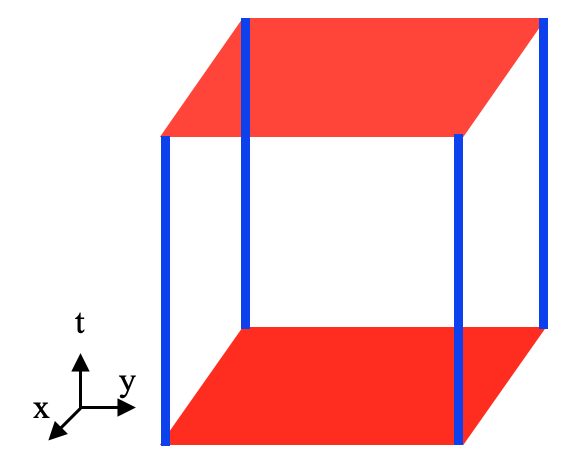}
    \caption{In the unit cell of spacetime cubic lattice, the time direction component of background gauge field $B^t$ lives on the blue vertical links and the spacial direction component of background gauge field $B^{xy}$ lives on the red horizontal face.}
    \label{fig:gaugefield}
\end{figure}

In summary, as shown in Figure \ref{fig:gaugefield}, the spacial direction component $B^{xy}$ lives on the face of horizontal plaquettes and the time direction component $B^t$ lives on the vertical links. Turning on background gauge fields is equivalent to inserting the defects line operators, where the lines cross the face or link with nontrivial gauge configurations. Since the line operators are closed, we also arrive at the flatness condition: $\Delta_t B^{xy} - \Delta_x\Delta_y B^t=0$.

\paragraph{Partition functions with gauge fields:}
Let us denote the partition function in terms of gauge fields as $Z_{\text{bos}}(B^t, B^{xy})$:
\begin{eqnarray}\label{ZB3d}
Z_{\text{bos}}(B^t, B^{xy}):= \Tr_{\CH_{\ft}} \left(\prod_{i=1}^{L_x} (U^y_i)^{W^B_{t,y;i}} \right) \left(\prod_{j=1}^{L_y} (U^x_j)^{W^B_{t,x;j}} \right)e^{-\beta H},
\end{eqnarray}
where $\CH_{\ft}$ is the twisted Hilbert space specified by the boundary condition $\ft=\{ t^x_j, t^y_i, t^{xy}\}$, which are related to the holonomies of $B^{xy}$ via \eqref{WBt}. As we remarked above, the partition function is invariant under the global shift \eqref{globalshift} due to \eqref{trivialU}. 
Moreover, the partition function $Z_{\text{bos}}[\fu, \ft]$ in terms of the twist and \twine sectors is given by \eqref{Zbosut}. We emphasize that although the partition function formally depends on $\{t^x_j, t^y_i\}$, we require that they depend only on the $\{\widehat{t}^x_j, \widehat{t}^y_i\}$. This is true for all the examples we consider in this note. 

\paragraph{Relations between partition functions: }
We proceed to determine the relation between the partition functions in the two pictures. Combining \eqref{ZB3d} and \eqref{Zbosut}, we find 
\begin{align}\label{ZuBboson}
\begin{split}
    Z_{\text{bos}}[\fu, \ft]&= \Tr_{\CH_{\ft}} \left(\prod_{i=1}^{L_x}\frac{1+ u_i^y U_{i}^y}{2} \right)\left(\prod_{j=1}^{L_y}\frac{1+u_j^x U_{j}^x}{2} \right)e^{-\beta H_{\text{bos}}}\\
    &= \frac{1}{2^{L_x+L_y}} \sum_{W^B_{t,x;j}, W^B_{t,y;i}=0,1 } \Tr_{\CH_{\ft}}   \left(\prod_{i=1}^{L_x} (u^y_i U^y_i)^{W^B_{t,y;i}}\right)\left( \prod_{j=1}^{L_y} (u^x_j U^x_j)^{W^B_{t,x;j}}\right) e^{-\beta H_{\text{bos}}}\\
    &= \frac{1}{2^{L_x+L_y}}\sum_{W^B_{t,x;j}, W^B_{t,y;i}=0,1 }\left(\prod_{i=1}^{L_x} (u^y_i )^{W^B_{t,y;i}}\right)\left( \prod_{j=1}^{L_y} (u^x_j )^{W^B_{t,x;j}}\right) Z_{\text{bos}}(B^t, B^{xy}).
\end{split}
\end{align}
As we see, it is essentially a discrete Fourier transformation in terms of the conjugating variables $\fu$ and $W^B_{t, \cdot;\cdot}$, subjected to constraints. The constraint on the left hand side is $\prod_{i=1}^{L_x} u_{i}^y \prod_{j=1}^{L_y} u_{j}^x=1$. This constraint is reproduced from the right hand side by noting that the right hand side should be invariant under the global shift \eqref{globalshift}, as expected. The $\ft$ on the left hand side is determined from the right hand side via \eqref{WBt}.  One can further work out the inverse relation, 
\begin{align}\label{ZBuboson}
\begin{split}
    Z_{\text{bos}}(B^t, B^{xy})&= \Tr_{\CH_{\ft}} \left(\prod_{i=1}^{L_x} (U^y_i)^{W^B_{t,y;i}} \right) \left(\prod_{j=1}^{L_y} (U^x_j)^{W^B_{t,x;j}} \right)e^{-\beta H}\\
    &=\sum_{u_i^y, u_j^x=\pm 1}\Tr_{\CH_{\ft}} \left(\prod_{i=1}^{L_x} (u_i^y)^{W^B_{t,y;i}} \frac{1+u_i^y U_i^y}{2}\right)\left(\prod_{j=1}^{L_y} (u_j^x)^{W^B_{t,x;j}} \frac{1+u_j^x U_j^x}{2}\right)e^{-\beta H_{\text{bos}}}\\
    &=\sum_{u_i^y, u_j^x=\pm 1} \left(\prod_{i=1}^{L_x} (u_i^y)^{W^B_{t,y;i}}\right)\left(\prod_{j=1}^{L_y} (u_j^x)^{W^B_{t,x;j}}\right) Z_{\text{bos}}[\fu, \ft].
\end{split}
\end{align}
Again, the gauge redundancy \eqref{globalshift} on the left hand side is reproduced from the right by the constraint \eqref{trivialU}. 

\subsubsection{Gauge fields for fermion systems}

Having established the relation between two presentations of the partition function of the bosonic theory, we can further determine the analogous relation for the fermionic theory.  Denoting the background fields for the $\Z_2^F$ subsystem fermion parity symmetry as $A^t, A^{xy}$, we have
\begin{align}\label{ZuAfermion}
\begin{split}
    Z_{\text{fer}}[\fu_f, \ft_f]= \frac{1}{2^{L_x+L_y}}\sum_{W^A_{t,x;j}, W^A_{t,y;i}=0,1 }\left(\prod_{i=1}^{L_x} (u^y_{f,i} )^{W^A_{t,y;i}}\right)\left( \prod_{j=1}^{L_y} (u^x_{f,j} )^{W^A_{t,x;j}}\right) Z_{\text{fer}}(A^t, A^{xy}),
\end{split}
\end{align}
and the inverse relation
\begin{eqnarray}\label{ZAufermion}
Z_{\text{fer}}(A^t, A^{xy})=\sum_{u_{f,i}^y, u_{f,j}^x=\pm 1} \left(\prod_{i=1}^{L_x} (u_{f,i}^y)^{W^A_{t,y;i}}\right)\left(\prod_{j=1}^{L_y} (u_{f,j}^x)^{W^A_{t,x;j}}\right) Z_{\text{fer}}[\fu_f, \ft_f].
\end{eqnarray}

\subsection{Boson-fermion duality with gauge fields}

In this subsection, we recast the boson-fermion duality in terms of the twist and \twine sectors as the boson-fermion duality coupled to gauge fields. Along the way, we find the subsystem generalization of the Arf invariant in $(2+1)$d. 

We start with the partition function of the $\Z_2^F$ subsystem-symmetric fermionic theory coupled to the background gauge field $A^t, A^{xy}$, and derive its relation with the bosonic counterpart. This relation generalizes \eqref{bosfer2d}  in $(1+1)$d. We start with \eqref{ZAufermion},  use \eqref{3dduality1} as well as \eqref{3dduality2}, and finally use \eqref{ZuBboson}, 
\begin{align}\label{331}
\begin{split}
    Z_{\text{fer}}(A^t, A^{xy})&=\sum_{u_{f,i}^y, u_{f,j}^x=\pm 1} \left(\prod_{i=1}^{L_x} (u_{f,i}^y)^{W^A_{t,y;i}}\right)\left(\prod_{j=1}^{L_y} (u_{f,j}^x)^{W^A_{t,x;j}}\right) Z_{\text{fer}}[\fu_f, \ft_f]\\
    &=\sum_{u_{i}^y, u_{j}^x=\pm 1} \left(\prod_{i=1}^{L_x} (u_{i}^y)^{W^A_{t,y;i}}\right)\left(\prod_{j=1}^{L_y} (u_{j}^x)^{W^A_{t,x;j}}\right) Z_{\text{bos}}[\fu, \ft]\\
    &=\frac{1}{2^{L_x+L_y}} \sum_{\substack{u_i^y, u_j^x=\pm 1\\W^b_{t,x;j}, W^b_{t,y;i}=0,1}}\left( \prod_{i=1}^{L_x} (u^y_i)^{W^A_{t,y;i}+W^b_{t,y;i}}\right) \left( \prod_{j=1}^{L_y} (u^x_j)^{W^A_{t,x;j}+W^b_{t,x;j}}\right)Z_{\text{bos}}(b^t, b^{xy}),
\end{split}
\end{align}
where $\ft$ and 
$\ft_{f}$ are related to the holonomies of the gauge fields via \eqref{WBt}, \eqref{WBtc} and $ \widehat{t}^y_{f,i}= (-1)^{W^A_{y;i}}, \widehat{t}^x_{f,j}= (-1)^{W^A_{x;j}}, t^{xy}_{f}= \prod_{i=1}^{L_x}(-1)^{W^A_{y;i}}= \prod_{j=1}^{L_y}(-1)^{W^A_{x;j}}$. The summation of $u$'s is subjected to the constraint $\prod_{i=1}^{L_x} u^y_i \prod_{j=1}^{L_y} u^x_j=1$.  Using the duality relations \eqref{3dduality1}, we find 
\begin{eqnarray}\label{332}
u_j^x u_{j+1}^x= (-1)^{W^A_{x;j}+ W^b_{x;j}}, \hspace{1cm} 1=(-1)^{W^A_{y;i}+ W^b_{y;i}}, \hspace{1cm} \forall ~i,j.
\end{eqnarray}
Substituting  \eqref{332} into the last line of  \eqref{331}, we find that, schematically,  the summation of $u$'s is replaced by the summation of the spatial holonomies of the gauge field $b$. Relegating the details of calculation to Appendix \ref{app:3ddualityproof},  the final result is that the boson-fermion duality in \eqref{3dduality2} can be recast as
\begin{equation}\label{3ddualitygauge}
Z_{\text{fer}}(A^t, A^{xy}) =\frac{1}{2^{L_y}} \sum_{\substack{W^b_{t,L_x,j}=0,1\\ W^b_{x,j}=0,1}}\left(\prod_{j=1}^{L_y-1} (-1)^{(W^A_{x,j}+ W^b_{x,j})\sum_{j'=1}^{j}( W^A_{t,L_x,j'}+W^b_{t,L_x,j'})}\right) Z_{\text{bos}}(b^t, b^{xy}),
\end{equation}
subjected to the constraint 
\begin{align}\label{3ddualityconstraints}
\begin{split}
    W^A_{t,y;i}+W^b_{t,y;i}+ W^A_{t,y;L_x}+ W^b_{t,y;L_x}&=0\mod 2, \hspace{1cm} i=1, ..., L_x-1,\\
    \sum_{j=1}^{L_y}(W^A_{t;L_x,j}+W^b_{t;L_x,j})&=0\mod 2,\\
    W^A_{y;i}+ W^b_{y;i}&=0 \mod 2 , \hspace{1cm} i=1, ..., L_x.
\end{split}
\end{align}
Let us make some comments on the boson-fermion duality \eqref{3ddualitygauge}.
\begin{enumerate}
    \item The boson-fermion duality is reminiscent to that in $(1+1)$d discussed previously in \eqref{bosfer2d}. A new feature is that the (holonomies of) dynamical gauge field $b$ is now subjected to nontrivial constraints \eqref{3ddualityconstraints}. 
    \item The two spatial directions are not on equal footing. In particular, the pre-factor in front of $Z_{\text{bos}}$ does not depend on the spatial holonomy of $b$ along the $y$ direction, i.e. $W^b_{y;i}$. Instead, it is completely fixed by the constraint \eqref{3ddualityconstraints}. The asymmetry
    between the two directions is the consequence of the JW duality which in itself is asymmetric by definition: the tail dressing the fermion only winds in the $x$ direction as shown in Figure \ref{fig:JWx}.
    \item Comparing with \eqref{bosfer2d}, it is tempting to define the pre-factor in front of $Z_{\text{bos}}$ in \eqref{3ddualitygauge} as the $(2+1)$d analogue of the Arf invariant, which we call the ``subsystem Arf invariant'' and denote by $\Arf_{L_x, L_y}(A^t+b^t, A^{xy}+b^{xy})$. Hence the duality \eqref{3ddualitygauge} schematically takes the form  
    \begin{eqnarray}
    \text{Fer}= \frac{\text{Bos}\times \text{(Subsystem Arf)}}{\Z_2},
    \end{eqnarray}
    which is similar to \eqref{2dferboskitaev}. 
    In section \ref{sec:3darf}, we discuss the properties of this $(2+1)$d Arf invariant in more detail, and show that it has the foliation structure along the $y$ direction, where each 2d layer is the standard $(1+1)$d Arf invariant \eqref{2darf}. 
\end{enumerate}

\section{\texorpdfstring{$(2+1)$d subsystem Arf invariant with $\Z_2^F$ subsystem fermion parity symmetry}{(2+1)d subsystem Arf invariant with Z2F subsystem fermion parity symmetry}}
\label{sec:3darf}

Comparing the standard $(1+1)$d boson-fermion duality \eqref{bosfer2d} and the $(2+1)$d boson-fermion duality with subsystem symmetry we found in \eqref{3ddualitygauge} (with constraints \eqref{3ddualityconstraints}), we are motivated to define the $(2+1)$d subsystem Arf invariant as 
\begin{eqnarray}\label{3darf}
\Arf_{L_x, L_y}(A^t, A^{xy}):= \sum_{j=1}^{L_y-1} \left( W^A_{x;j} \sum_{j'=1}^{j} W^A_{t;L_x,j'}\right),
\end{eqnarray}
subjected to the constraints\footnote{\label{footnote1}One can solve the third constraint by defining another gauge field $A'$ such that $A^t_{k,i,j}=A'^{t}_{k,i,j},A^{xy}_{k,i,j}= A'^{x}_{k,i,j}-A'^{x}_{k,i,j-1}$. The new gauge field $A'$ is a flat gauge field satisfying $\Delta_x A'^{t}-\Delta_tA'^{x}=0$ with constraints $W^{A'}_{t,y;i}=W^{A'}_{t,y;L_x}$, $\sum_{j=1}^{L_y}W_{t;L_x,j}^{A'}=0$.  
In terms of $A'$, the subsystem Arf invariant can be simplified as
\begin{align}
\begin{split}
    \Arf_{L_x,L_y}(A'^t,A'^{x})&= \sum_{j=1}^{L_y-1} \left( W_{x;j}^A \sum_{j'=1}^{j} W_{t;L_x,j'}^{A}\right)= \sum_{j=1}^{L_y} \left( W_{x;j}^A \sum_{j'=1}^{j} W_{t;L_x,j'}^{A}\right)=\sum_{j=1}^{L_y}\left( W_{t;L_x,j}^A \sum_{j'=j}^{L_y} W_{x;j'}^{A}\right)\nonumber\\
    &= \sum_{j=1}^{L_y}\left( W_{t;L_x,j}^{A'} (W_{x;L_y}^{A'}- W_{x;j-1}^{A'})\right)
    =  \sum_{j=1}^{L_y}W^{A'}_{t;L_y, j}W^{A'}_{x;j-1}.
\end{split}
\end{align}
where $j\sim j+L_y$. 
In terms of $A'$, we find that the foliation structure is more transparent. However, we emphasize that this is not a trivial stacking of $(1+1)$d Arf invariants because of the constraints above. We thank Ho Tat Lam who informed us about this simplification. }
\begin{align}\label{3dconstr}
\begin{split}
    W^A_{t,y;i}+ W^A_{t,y;L_x}&=0\mod 2, \hspace{1cm} i=1, ..., L_x-1,\\
    \sum_{j=1}^{L_y}W^A_{t;L_x,j}&=0\mod 2,\\
    W^A_{y;i}&=0 \mod 2 , \hspace{1cm} i=1, ..., L_x.
\end{split}
\end{align}
Let us comment on its properties. 
\begin{enumerate}
    \item Similar to the $(1+1)$d case in Section \ref{sec.arfJW}, we emphasize that the Arf invariant can not be written as an integral of a local Lagrangian density. Instead, it depends on the gauge field via its holonomies.
    \item It is asymmetric in the time direction, the spatial $x$ direction and the spatial $y$ direction. In fact, as we will see, there is a foliation structure along the spatial $y$ direction, but not in the other two. 
    \item The first constraint in \eqref{3dconstr} shows that the gauge choice $L_x$ made in the calculation is not special, i.e.\  $W^A_{t,y;i}$ are equivalent mod 2 for all $i$. Hence both \eqref{3darf} and \eqref{3dconstr} are homogeneous in the $x$ diction.  However, the action \eqref{3darf} and the constraint \eqref{3dconstr} are non-homogeneous within the $y$ direction: 
    $j=L_y$ contribution in the action \eqref{3darf} is missing due to the second constraint in \eqref{3dconstr}. This non-homogeneity is an obstruction to taking the continuum limit, and we will leave the discussion of the continuum version of our duality in future work. This subtlety should be understood as the peculiar feature of the subsystem symmetry, which requires more complicated structures than trivially stacking the theories along the $y$ direction.
    \item There are eight other variations of the 3d Arf invariant, related to \eqref{3darf} and \eqref{3dconstr} via $D_4$ transformation. See figure \ref{fig:8JW} for the relation between eight variations. 
\end{enumerate}

\subsection{Foliation structure}

We would like to show that the subsystem Arf invariant enjoys the foliation structure (in the sense we specify below), as expected for generic models with subsystem symmetry or with the fractonic behavior. To see this, let us consider the 3d Arf invariant on a spatial lattice with size $L_x\times (L_y+1)$, 
\begin{eqnarray}\label{3darf2}
\Arf_{L_x, L_y+1}(A^t, A^{xy})= \sum_{j=1}^{L_y} \left( W^A_{x;j} \sum_{j'=1}^{j} W^A_{t;L_x,j'}\right),
\end{eqnarray}
with the constraint 
\begin{align}\label{3dconst2}
\begin{split}
    W^A_{t,y;i}+ W^A_{t,y;L_x}&=0\mod 2, \hspace{1cm} i=1, ..., L_x-1,\\
    \sum_{j=1}^{L_y+1}W^A_{t;L_x,j}&=0\mod 2,\\
    W^A_{y;i}&=0 \mod 2 , \hspace{1cm} i=1, ..., L_x.
\end{split}
\end{align}
To compare with the 3d subsystem Arf invariant on $L_x\times L_y$ lattice, we reorganize the 3d subsystem Arf invariant \eqref{3darf2} on the $L_x\times (L_y+1)$ lattice as
\begin{align}\label{3darf3}
\begin{split}
    \Arf_{L_x, L_y+1}(A^t, A^{xy})&= \sum_{j=1}^{L_y-1} \left( W^A_{x,j} \sum_{j'=1}^{j} W^A_{t;L_x,j'}\right) + W^A_{x;L_y} \sum_{j=1}^{L_y} W^A_{t;L_x,j}.
\end{split}
\end{align}
The first factor on the right hand side of \eqref{3darf3} has the same form as the 3d subsystem Arf invariant on the $L_x\times L_y$ lattice, hence we label it as $\widetilde{\Arf}_{L_x,L_y}(A^t, A^{xy})$. We used the tilde to emphasize that the gauge field satisfies the constraints in the $L_x\times (L_y+1)$ system \eqref{3dconst2} rather than those in the $L_x\times L_y$ system \eqref{3dconstr}. Moreover, we can use the second contraint in \eqref{3dconst2} to rewrite  the second factor in \eqref{3darf3} as $W^A_{x;L_y} \sum_{j=1}^{L_y} W^A_{t;L_x,j}= W^A_{x;L_y} W^A_{t;L_x,L_{y}+1}$, which is the holonomy in the spacial direction time a holonomy in the time direction, i.e.\ the standard $(1+1)$d Arf invariant. In summary, \eqref{3darf3} can be simplified to 
\begin{eqnarray}\label{3darf4}
\Arf_{L_x, L_y+1}(A^t, A^{xy})&= \widetilde{\Arf}_{L_x,L_y}(A^t, A^{xy}) + W^A_{x;L_y} W^A_{t;L_x,L_{y}+1}.
\end{eqnarray}
Eq.~\eqref{3darf4} shows the foliation structure of the 3d subsystem Arf invariant in the following sense: the 3d subsystem Arf invariant on system size $L_x\times (L_y+1)$ is equivalent to the stacking of another 3d subsystem Arf invariant (with a slightly modified constraint) on system size $L_x\times L_y$ and the standard 2d Arf invariant. 

\section{Applications} \label{sec:app3d}

We provide two applications of the general discussion in the previous section. We first apply the generalized JW transformation to show that  (the fermionization of) the plaquette Ising model is dual to the plaquette fermion model, which is a generalization of the duality Ising $\leftrightarrow$ Majorana fermion in $(1+1)$d. A key difference is that both sides of the duality are interacting theories and do not seem to be exactly solvable. We further discuss a duality between two fermion theories, and show that the quartic fermion model has an exactly vanishing partition function, although the model is not exactly solvable. 

\subsection{\texorpdfstring{Exact duality: plaquette Ising $\leftrightarrow$ plaquette fermion}{Exact duality: plaquette Ising <-> plaquette fermion}}
\label{app:3d1}

We have established the boson-fermion duality between a generic bosonic system with a $\Z_2$ subsystem symmetry and a generic fermion system with a $\Z_2^F$ subsystem fermion parity symmetry. We now apply the duality to a concrete model, the plaquette Ising model in Section \ref{app:3d1}, and show that it is dual to the plaquette fermion model, with symmetry sectors properly exchanged according to \eqref{3dduality1} (or equivalently coupling to gauge fields according to \eqref{3ddualitygauge}). 

The Hamiltonian of the plaquette Ising model is 
\begin{align}\label{PlaIsingHam}
\begin{split}
H_{\text{PIsing}}=&-\sum_{i=1}^{L_x-1}\sum_{j=1}^{L_y-1} Z_{i,j} Z_{i+1,j} Z_{i,j+1} Z_{i+1,j+1} -\sum_{i=1}^{L_x}\sum_{j=1}^{L_y} X_{i,j} \\
&- \sum_{i=1}^{L_x-1} 
t_i^yt_{i+1}^y Z_{i,L_y}Z_{i+1,L_y}Z_{i,1}Z_{i+1,1} - \sum_{j=1}^{L_y-1} 
t_j^xt_{j+1}^x Z_{L_x,j}Z_{1,j}Z_{L_x,j+1}Z_{1,j+1}\\
&-
t_1^xt_1^y t_{L_y}^x t_{L_x}^y t^{xy} Z_{L_x,L_y}Z_{1,L_y}Z_{L_x,1}Z_{1,1}.
\end{split}
\end{align}
This Hamiltonian preserves the symmetry exchanging the two spatial directions $x\leftrightarrow y$.  
Note that the Hamiltonian depends on the twist parameters $\ft=\{t^x_j, t^y_i, t^{xy}\}$ only through their combinations $\{\widehat{t}^x_j, \widehat{t}^y_i\}$, i.e.
\begin{align}\label{PlaIsing2}
\begin{split}
H_{\text{PIsing}}=&-\sum_{i=1}^{L_x-1}\sum_{j=1}^{L_y-1} Z_{i,j} Z_{i+1,j} Z_{i,j+1} Z_{i+1,j+1} -\sum_{i=1}^{L_x}\sum_{j=1}^{L_y} X_{i,j} \\
&- \sum_{i=1}^{L_x-1} 
\widehat{t}_i^y Z_{i,L_y}Z_{i+1,L_y}Z_{i,1}Z_{i+1,1} - \sum_{j=1}^{L_y-1} 
\widehat{t}_j^x Z_{L_x,j}Z_{1,j}Z_{L_x,j+1}Z_{1,j+1}\\
&-
\widehat{t}_{L_y}^x \widehat{t}_{L_x}^y \left(\prod_{j=1}^{L_y} \widehat{t}^x_j\right) Z_{L_x,L_y}Z_{1,L_y}Z_{L_x,1}Z_{1,1}.
\end{split}
\end{align}
The Hamiltonian is obviously $\Z_2$ subsystem-symmetric: the transformation $Z_{i,j}\to -Z_{i,j}$ for all $i=1, ..., L_x$ and an arbitrary fixed $j$ (or a similar transformation with the role of $i, j$ exchanged) leaves the Hamiltonian invariant.  

By using the generalized JW map \eqref{JWxmaj}, the plaquette Ising model can be rewritten in terms of the real fermion operators,  and we get the plaquette fermion model with suitable boundary conditions
\begin{align}\label{PlaFerHam}
\begin{split}
H_{\text{Pfer}}=&\sum_{i=1}^{L_x-1}\sum_{j=1}^{L_y-1} \gamma'_{i,j} \gamma_{i+1,j} \gamma'_{i,j+1} \gamma_{i+1,j+1} +i\sum_{i=1}^{L_x}\sum_{j=1}^{L_y} \gamma_{i,j}\gamma'_{i,j} \\
&+ \sum_{i=1}^{L_x-1} 
t_i^yt_{i+1}^y\gamma'_{i,L_y}\gamma_{i+1,L_y}\gamma'_{i,1}\gamma_{i+1,1} + \sum_{j=1}^{L_y-1} 
t_j^xt_{j+1}^x P_{f,j}^x P_{f,j+1}^x \gamma'_{L_x,j}\gamma_{1,j}\gamma'_{L_x,j+1}\gamma_{1,j+1}\\
&+
t_1^xt_1^y t_{L_y}^x t_{L_x}^y t^{xy}P_{f,L_y}^x P_{f,1}^x\gamma'_{L_x,L_y}\gamma_{1,L_y}\gamma'_{L_x,1}\gamma_{1,1}.
\end{split}
\end{align}
Further replacing the bilinears of $t$ by $\widehat{t}$'s, and using the condition $u_{f,i}^y=u_i^y, u_{f,j}^x = u_j^x$ where $u_{f,\cdot}^{\cdot}$ is the eigenvalue of $P_{f,\cdot}^{\cdot}$, we get
\begin{align}\label{PlaFerHam2}
\begin{split}
H_{\text{Pfer}}=&\sum_{i=1}^{L_x-1}\sum_{j=1}^{L_y-1} \gamma'_{i,j} \gamma_{i+1,j} \gamma'_{i,j+1} \gamma_{i+1,j+1} +i\sum_{i=1}^{L_x}\sum_{j=1}^{L_y} \gamma_{i,j}\gamma'_{i,j} \\
&+ \sum_{i=1}^{L_x-1} 
\widehat{t}_i^y\gamma'_{i,L_y}\gamma_{i+1,L_y}\gamma'_{i,1}\gamma_{i+1,1} + \sum_{j=1}^{L_y-1} 
\widehat{t}_j^x u_{j}^x u_{j+1}^x \gamma'_{L_x,j}\gamma_{1,j}\gamma'_{L_x,j+1}\gamma_{1,j+1}\\
&+
\widehat{t}_{L_y}^x \widehat{t}_{L_x}^y \left(\prod_{j=1}^{L_y} \widehat{t}^x_j\right)u_{L_y}^x u_{1}^x\gamma'_{L_x,L_y}\gamma_{1,L_y}\gamma'_{L_x,1}\gamma_{1,1}.
\end{split}
\end{align}
Let us make some comments about the fermionic theory \eqref{PlaFerHam} and \eqref{PlaFerHam2}:
\begin{enumerate}
    \item Since the interactions among the fermions are supported on the plaquettes, rather than on the links, we call this model the plaquette fermion model. 
    \item The plaquette fermion model has a $\Z_2^F$ subsystem fermionic symmetry: the fermion parity transformation $\gamma_{i,j}\to -\gamma_{i,j}$ and $\gamma'_{i,j}\to -\gamma'_{i,j}$ for all $i=1, ..., L_x$ and an arbitrary fixed $j$ (or the same transformation with the role of $i$ and $j$ exchanged) leaves the theory invariant. Moreover, the quadratic fermion interactions across the links, e.g. $\gamma_{i, j} \gamma_{i, j+1}$, are forbidden by the $\Z_2^F$ subsystem fermion parity symmetry. 
    \item Although the plaquette Ising model is invariant under the exchange of the two spatial directions, the plaquette fermion model is not. The asymmetry between the two spatial directions originates from the asymmetry of the JW transformation. 
    \item A crucial difference here from $(1+1)$d is that although the fermion model in $(1+1)$d is quadratic (hence exactly solvable), the plaquette fermion model is not. The latter contains four-fermion interactions and does not appear to be exactly solvable. 
    \item One can use other versions of the JW map (exhibited in Figure \ref{fig:8JW}) to find other fermionic dual descriptions of the plaquette Ising model. Hence we can establish a web of dualities (a multi-ality). We will not attempt to build this web explicitly.
    \item The continuum field theory description of the plaquette Ising model (and its generalization) was studied in \cite{Seiberg:2019vrp, Pretko:2018jbi}. It would be interesting to study the continuum limit of the plaquette fermion model as well. If it is possible, we would expect to find an infrared duality between the two continuum field theories induced by the exact duality between \eqref{PlaIsing2} and \eqref{PlaFerHam2}, generalizing the discussions in Section \ref{sec:2dcont}. 
\end{enumerate}

\subsection{Exactly vanishing partition function of the quartic fermion theory}
\label{sec:quarticfermion}

In this subsection, we consider another fermionic model with quartic interactions, which modifies the plaquette fermion model \eqref{PlaFerHam} discussed in section \ref{app:3d1}. By following an analogous discussion in section \ref{sec:2dexact}, we show that the partition function of the quartic fermion model vanishes exactly under certain boundary conditions, despite that the fermion model is not exactly solvable.\footnote{We thank Yuan Yao for suggesting this problem to us. }

The Hamiltonian for the quartic fermion model is 
\begin{equation}\label{qfer}
\begin{split}
    H_{\text{Qfer}}= &\sum_{i=1}^{L_x-1}\sum_{j=1}^{L_y} \gamma'_{i,j} \gamma_{i+1,j} \gamma'_{i,j+1} \gamma_{i+1,j+1} +\sum_{i=1}^{L_x}\sum_{j=1}^{L_y} \gamma_{i,j}\gamma'_{i,j}\gamma_{i,j+1}\gamma'_{i,j+1} \\
    & + \sum_{j=1}^{L_y} 
t_{f,j}^xt_{f,j+1}^x \gamma'_{L_x,j}\gamma_{1,j}\gamma'_{L_x,j+1}\gamma_{1,j+1}.
\end{split}
\end{equation}
This Hamiltonian differs from the plaquette fermion model \eqref{PlaFerHam} by replacing the quadratic on-site term with the quartic on-site term. 
Note that for simplicity we do not activate the twisted boundary conditions along the $y$ direction, by assuming $t^y_{f,i}=1$ for all $i$. We also only consider the fermion parity subsystem symmetry along the $x$ direction, whose generators are
\begin{eqnarray}
P_{f,j}^{x}= \prod_{i=1}^{L_x} \left( -i \gamma_{i,j} \gamma_{i,j}'\right), \hspace{1cm} j=1, ..., L_x.
\end{eqnarray}
Let us again consider a map between two sets of fermion operators,
\begin{equation}\label{fftrans3d}
\begin{array}{rlrl}
    \widetilde{\gamma}_{i,j}'&= \gamma_{i+1,j},& \hspace{0cm} i&=1, ..., L_x-1,~~ j=1, ..., L_y,\\
    \widetilde{\gamma}'_{L_x, j}&= -t_f \gamma_{1,j},&\hspace{0cm} j&=1, ..., L_y,\\
    \widetilde{\gamma}_{i,j}&= \gamma_{i,j}', &\hspace{0cm} i&=1, ..., L_x,~~ j=1,..., L_y ,
\end{array}
\end{equation}
which generalizes \eqref{fftrans2d}. Applying \eqref{fftrans3d}, the quartic hopping terms and the on-site interaction terms in the quartic fermion model \eqref{qfer} (in terms of $\gamma$'s) are exchanged, and we get another quartic fermion model (in terms of $\widetilde{\gamma}$'s) which happen to have the same Hamiltonian 
\begin{align}
\begin{split}
    H_{\widetilde{\text{Qfer}}}= &\sum_{i=1}^{L_x-1}\sum_{j=1}^{L_y} \widetilde{\gamma}'_{i,j} \widetilde{\gamma}_{i+1,j} \widetilde{\gamma}'_{i,j+1} \widetilde{\gamma}_{i+1,j+1} +i\sum_{i=1}^{L_x}\sum_{j=1}^{L_y} \widetilde{\gamma}_{i,j}\widetilde{\gamma}'_{i,j}\widetilde{\gamma}_{i,j+1}\widetilde{\gamma}'_{i,j+1} \\
    & \qquad \qquad + \sum_{j=1}^{L_y} 
t_{f,j}^xt_{f,j+1}^x \widetilde{\gamma}'_{L_x,j}\widetilde{\gamma}_{1,j}\widetilde{\gamma}'_{L_x,j+1}\widetilde{\gamma}_{1,j+1}.
\end{split}
\end{align}
The subsystem fermion parity operator in terms of the new fermions is 
\begin{eqnarray}
\widetilde{P}_{f,j}^x= \prod_{i=1}^{L_x} \left( -i \widetilde{\gamma}_{i,j} \widetilde{\gamma}_{i,j}'\right) = t_{f,j}^x P_{f,j}^x, \hspace{1cm} j=1, ..., L_x.
\end{eqnarray}
As the Hamiltonians happen to have the same form, we have the relation between the partition functions
\begin{eqnarray}\label{3dzero}
Z_{\text{Qfer}}[u_{f,j}^x, t_{f,j}^x]= Z_{\widetilde{\text{Qfer}}}[u_{f,j}^x t_{f,j}^x, t_{f,j}^x]\equiv Z_{{\text{Qfer}}}[u_{f,j}^x t_{f,j}^x, t_{f,j}^x].
\end{eqnarray}
Following similar discussion in Section \ref{sec:2dexact}, this means that the partition function exactly vanishes as long as there is at least one $j=1, ..., L_y$ such that the $j$-th $(t,x)$-layer along the $y$ direction  has the RR spin structure. The above discussion does not apply to the plaquette fermion model \eqref{PlaFerHam} due to the asymmetry between the hopping term and the on-site interaction. 

We finally comment that the transformation \eqref{fftrans3d} should be the composition of JW transformation, followed by a KW transformation (which we have not discussed in this paper), and another JW transformation associated to the $\Z_2$ subsystem symmetry, similar to the situation in $(1+1)$d discussed in Section \ref{sec:2dexact}.  

\section{Comments on future directions}
\label{sec:comments}

In this paper, we find the subsystem Arf invariant for $(2+1)$d models with subsystem symmetry. Starting from this, there are many interesting directions to explore.
\begin{enumerate}
    \item It would be interesting to perform a parallel analysis for the exact boson-boson duality with $\Z_2$ subsystem symmetry induced by the KW transformation. Note that the KW transformation amounts to gauging the $\Z_2$ subsystem symmetry. As there exists $\Z_2$ (strong and weak) subsystem symmetry protected topological (SSPT) states in $(2+1)$d \cite{You:2018oai,Burnell:2021reh}, one can discuss the twisted KW transformation by first stacking an SSPT and then gauging the subsystem symmetry. Such twisted KW transformation would significantly enrich the duality web.
    \item Another interesting direction is to generalize the discussion in this work to $(3+1)$d. The $\Z_2$ subsystem symmetry can be straightforwardly generalized to $(3+1)$d, and the $\Z_2$ strong and weak SSPTs have been studied in \cite{You:2018oai, Devakul:2019duj,Burnell:2021reh}. It would be interesting to explore the web of boson-boson/boson-fermion/fermion-fermion dualities induced by twisted KW transformation or twisted JW transformation or their combinations, and find dynamical applications.
    \item Similar to $(1+1)$d, one can study the boundary anomaly of the subsystem Arf invariant or a subsystem analogue of the Kitaev chain which captures this subsystem Arf invariant. In $(1+1)$d, the nontrivial phase of the Kitaev model (whose partition function is the Arf invaraint) is mapped to the $\Z_2$ symmetry breaking phase of the Ising model. Similarly, the fermionic exactly solvable model whose partition function is the subsystem Arf invariant is mapped to the $\Z_2$ subsystem symmetry breaking phase of the plaquette Ising model. We thus find the exactly solvable Hamiltonian associated to the subsystem Arf invariant to be
    \begin{equation}
    \begin{split}
    H_{\text{SubArf}}= &\sum_{i=1}^{L_x-1}\sum_{j=1}^{L_y-1} \gamma'_{i,j} \gamma_{i+1,j} \gamma'_{i,j+1} \gamma_{i+1,j+1}+ \sum_{i=1}^{L_x-1} 
\widehat{t}_i^y\gamma'_{i,L_y}\gamma_{i+1,L_y}\gamma'_{i,1}\gamma_{i+1,1} \\&+ \sum_{j=1}^{L_y-1} 
\widehat{t}_j^x u_{j}^x u_{j+1}^x \gamma'_{L_x,j}\gamma_{1,j}\gamma'_{L_x,j+1}\gamma_{1,j+1}+
\widehat{t}_{L_y}^x \widehat{t}_{L_x}^y \left(\prod_{j=1}^{L_y} \widehat{t}^x_j\right)u_{L_y}^x u_{1}^x\gamma'_{L_x,L_y}\gamma_{1,L_y}\gamma'_{L_x,1}\gamma_{1,1}.
\end{split}
    \end{equation}
    Different terms are mutually commuting. 
This model exhibits interesting boundary properties. When we impose the open boundary condition along the $x$ direction, there are fermions that do not enter the Hamiltonian, hence are dangling fermions. They form gapless boundary modes. Whereas if we impose the open boundary condition along the $y$ direction, there are no dangling fermions on the boundary, hence no boundary modes. This is as expected since as we have seen in Section \ref{sec:3darf}, the subsystem Arf invariant is roughly speaking a stacking of $(1+1)$d Arf invariant along the $y$ direction. It would be interesting to further explore the role of anomaly inflow into the $(1+1)$d system on the boundary with $\Z_2^F$ subsystem fermion parity anomaly. 
\end{enumerate}

\section*{Acknowledgements}

This work grew out of a study group held at Kavli IPMU, and we would like to thank Abhiram Kidambi, Jacob M. Leedom, Linhao Li and Masaki Oshikawa for taking part in the stimulating discussions in the early stages. We thank Yuan Yao for suggesting the problem discussed in section \ref{sec:quarticfermion}, and Ho Tat Lam for comments leading to the footnote \ref{footnote1} as well as the comments on subsystem symmetric exactly solvable Kitaev model in Section \ref{sec:comments}. 
We also thank 
Ho Tat Lam and Yuan Yao for useful discussions, and
Ho Tat Lam, Nathan Seiberg, Shu-Heng Shao and Juven Wang 
for comments on a draft. Y.Z. thanks Yoshiki Fukusumi and Yuji Tachikawa for collaboration on a related project on fermionization in $(1+1)$d \cite{Fukusumi:2021zme}. This work is partially supported by World Premier International Research Center Initiative (WPI) Initiative, MEXT, Japan at Kavli IPMU, the University of Tokyo. 
W.C.\ is also supported by the Global Science Graduate Course (GSGC) program of the University of Tokyo, 
and acknowledges support from JSPS KAKENHI Grants No.19H058101015 and No.22J21553.
M.Y.\ is also supported in part by the JSPS Grant-in-Aid for Scientific Research (17KK0087, 19K03820, 19H00689, 20H05860). The authors of this paper were ordered alphabetically.


\appendix

\section{\texorpdfstring{Derivation of \eqref{3ddualitygauge} and \eqref{3ddualityconstraints}}{Derivation of (3.35) and (3.36)}}
\label{app:3ddualityproof}

In this appendix, we provide details of deriving \eqref{3ddualitygauge} and the constraints \eqref{3ddualityconstraints}. We start with \eqref{331} and \eqref{332}, which are reproduced here for convenience:
\begin{equation}\label{A1}
Z_{\text{fer}}(A^t, A^{xy})=\frac{1}{2^{L_x+L_y}} \! \sum_{\substack{u_i^y, u_j^x=\pm 1\\W^b_{t,x;j}, W^b_{t,y;i}=0,1}} \! \left( \prod_{i=1}^{L_x} (u^y_i)^{W^A_{t,y;i}+W^b_{t,y;i}}\right) \left( \prod_{j=1}^{L_y} (u^x_j)^{W^A_{t,x;j}+W^b_{t,x;j}}\right)Z_{\text{bos}}(b^t, b^{xy}),
\end{equation}
which is subjected to the condition 
\begin{eqnarray}\label{A2}
u_j^x u_{j+1}^x= (-1)^{W^A_{x;j}+ W^b_{x;j}}, \hspace{1cm} 1=(-1)^{W^A_{y;i}+ W^b_{y;i}}, \hspace{1cm} \forall ~i,j, \hspace{1cm}\prod_{i=1}^{L_x} u^y_i \prod_{j=1}^{L_y} u^x_j=1.
\end{eqnarray}
Let us first use the constraint $\prod_{i=1}^{L_x} u^y_i \prod_{j=1}^{L_y} u^x_j=1$ to eliminate $u^y_{L_x}$ in \eqref{A1}. The relevant factor becomes
\begin{align}\label{A3}
\begin{split}
    &\left( \prod_{i=1}^{L_x} (u^y_i)^{W^A_{t,y;i}+W^b_{t,y;i}}\right) \left( \prod_{j=1}^{L_y} (u^x_j)^{W^A_{t,x;j}+W^b_{t,x;j}}\right)\\
    &= \left( \prod_{i=1}^{L_x-1} (u^y_i)^{W^A_{t,y;i}+W^b_{t,y;i} + W^A_{t,y;L_x}+ W^b_{t,y;L_x}}\right) \left( \prod_{j=1}^{L_y} (u^x_j)^{W^A_{t,x;j}+W^b_{t,x;j}+ W^A_{t,y;L_x}+ W^b_{t,y;L_x}}\right).
\end{split}
\end{align}
To make use of the first condition in \eqref{A2}, we regroup the second factor in \eqref{A3} as follows, 
\begin{align}\label{A4}
\begin{split}
    & \prod_{j=1}^{L_y} (u^x_j)^{W^A_{t,x;j}+W^b_{t,x;j}+ W^A_{t,y;L_x}+ W^b_{t,y;L_x}}\\
    &=\left(\prod_{j=1}^{L_y-1} (u_j^x u_{j+1}^x)^{\sum_{j'=1}^{j}( W^A_{t,x;j'}+W^b_{t,x;j'}+ W^A_{t,y;L_x}+ W^b_{t,y;L_x})}\right)  (u^x_{L_y})^{\sum_{j'=1}^{L_y}(W^A_{t,x;j'}+W^b_{t,x;j'}+ W^A_{t,y;L_x}+ W^b_{t,y;L_x})}.
\end{split}
\end{align}
Substituting \eqref{A4} into \eqref{A3} and further into \eqref{A1}, we find 
\begin{align}\label{A5}
\begin{split}
    Z_{\text{fer}}(A^t, A^{xy})&=\frac{1}{2^{L_x+L_y}} \sum_{\substack{u_i^y, u_j^x=\pm 1\\W^b_{t,x;j}, W^b_{t,y;i}=0,1\\ i\neq L_x}}\left( \prod_{i=1}^{L_x-1} (u^y_i)^{W^A_{t,y;i}+W^b_{t,y;i}+ W^A_{t,y;L_x}+ W^b_{t,y;L_x}}\right)\\&\hspace{1cm}\times \left(\prod_{j=1}^{L_y-1} (u_j^x u_{j+1}^x)^{\sum_{j'=1}^{j}( W^A_{t,x;j'}+W^b_{t,x;j'}+ W^A_{t,y;L_x}+ W^b_{t,y;L_x})}\right)  \\&\hspace{1cm}\times  (u^x_{L_y})^{\sum_{j'=1}^{L_y}(W^A_{t,x;j'}+W^b_{t,x;j'}+ W^A_{t,y;L_x}+ W^b_{t,y;L_x})}Z_{\text{bos}}(b^t, b^{xy})\\
    &=\frac{1}{2^{L_y}} \sum_{\substack{W^b_{t,x;j}, W^b_{t,y;i}=0,1\\ W^b_{x,j}=0,1}}\left(\prod_{j=1}^{L_y-1} (-1)^{(W^A_{x;j}+ W^b_{x;j})\sum_{j'=1}^{j}( W^A_{t,x;j'}+W^b_{t,x;j'}+ W^A_{t,y;L_x}+ W^b_{t,y;L_x})}\right) Z_{\text{bos}}(b^t, b^{xy}) ,
\end{split}
\end{align}
where in the second equality, we summed over $u_i^y$ for $i=1, ..., L_x-1$ and $u^x_{L_y}$ (since they are now independent), which produces the constraints 
\begin{align}\label{A6}
\begin{split}
    W^A_{t,y;i}+W^b_{t,y;i}+ W^A_{t,y;L_x}+ W^b_{t,y;L_x}&=0\mod 2, \hspace{1cm} i=1, ..., L_x-1,\\
    \sum_{j=1}^{L_y}(W^A_{t,x;j}+W^b_{t,x;j}+ W^A_{t,y;L_x}+ W^b_{t,y;L_x})&=0\mod 2,\\
    W^A_{y;i}+ W^b_{y;i}&=0 \mod 2 , \hspace{1cm} i=1, ..., L_x,
\end{split}
\end{align}
where the third constraint comes from \eqref{A2}. 
In the second equality of \eqref{A5},  we also used the first equality in \eqref{A2} to substitute $u_j^x u_{j+1}^x$ by $(-1)^{W^A_{x,j}+ W^b_{x,j}}$. We can simplify the last line of \eqref{A5} and also the constraints \eqref{A6} by making use of the definition  \eqref{WBtdef}, 
\begin{equation}\label{A7}
Z_{\text{fer}}(A^t, A^{xy}) =\frac{1}{2^{L_y}} \sum_{\substack{W^b_{t;L_x,j}=0,1\\ W^b_{x;j}=0,1}}\left(\prod_{j=1}^{L_y-1} (-1)^{(W^A_{x;j}+ W^b_{x;j})\sum_{j'=1}^{j}( W^A_{t;L_x,j'}+W^b_{t;L_x,j'})}\right) Z_{\text{bos}}(b^t, b^{xy}) ,
\end{equation}
and the constraints become
\begin{align}\label{A8}
\begin{split}
    W^A_{t,y;i}+W^b_{t,y;i}+ W^A_{t,y;L_x}+ W^b_{t,y;L_x}&=0\mod 2, \hspace{1cm} i=1, ..., L_x-1,\\
    \sum_{j=1}^{L_y}(W^A_{t;L_x,j}+W^b_{t;L_x,j})&=0\mod 2,\\
    W^A_{y;i}+ W^b_{y;i}&=0 \mod 2 , \hspace{1cm} i=1, ..., L_x.
\end{split}
\end{align}
In fact, we can substitute the second constraint in \eqref{A8} into \eqref{A7} to let the product in the summation to be from $j=1$ to $L_y$ (instead of $L_y-1$), which makes the expression more symmetric. However, the asymmetry still persists, for example, in the first constraint in \eqref{A8}, where the holonomies in time direction is independent of the coordinate $i$ in $x$ direction. This finishes the proof of \eqref{3ddualitygauge} and \eqref{3ddualityconstraints}. 


\bibliographystyle{ytphys}
\baselineskip=.95\baselineskip
\bibliography{bib}


\end{document}